\documentclass[useAMS,usenatbib]{mn2e}
\input{psfig}
\usepackage{rotating}
\usepackage{natbib}
\usepackage[dvips]{epsfig}
\usepackage{subfigure}
\usepackage{afterpage}

\title[OH masers in M82]{OH main line masers in the M82 starburst}
\author[Argo et al]{M.K.~Argo, A.~Pedlar, R.J.~Beswick, T.W.B.~Muxlow\\
University of Manchester, Jodrell~Bank Observatory, Macclesfield, Cheshire 
SK11~9DL\\}

\def\farcs  {\hbox{$.\!\!^{\prime\prime}$}}
\def\kms    {\ifmmode {\rm km\,s}^{-1} \else km\,s$^{-1}$\fi}

\def\rasec  {\hbox{$.\!\!^{\rm s}$}}
\def\degr   {\hbox{$^\circ$}}

\begin{document}
\maketitle

\begin{abstract} 
{
A study of the distribution of OH gas in the central region of
the nearby active starburst galaxy M82 has confirmed two previously known
bright masers and revealed several new main line masers.  Three of
these are seen only at 1665\,MHz, one is detected only at 1667\,MHz, while
the rest are detected in both lines.  Observations covering both the 1665
and 1667 MHz lines, conducted with both the Very Large Array (VLA) and the
Multi-Element Radio Linked Interferometer Network (MERLIN), have been used
to accurately measure the positions and velocities of these features.  This
has allowed a comparison with catalogued continuum features in the
starburst such as H{\sc ii} regions and supernova remnants, as well as known water
and satellite line OH masers.  Most of the main line masers appear to be associated with
known H{\sc ii} regions although the two detected only at 1665\,MHz are
seen along the same line of sight as known supernova remnants.
}
\end{abstract}



\section{Introduction}

Astronomical masers are useful probes of the interstellar medium in star
forming regions, both in the Milky Way and external galaxies.  Masers usually occur
under specific conditions found in star forming regions so the presence of 
a maser immediately provides information on the state of the ISM and acts 
as an indicator for ongoing star formation.

Studies of star forming regions within the Milky Way have the advantage 
that the masers can be studied at high linear resolution.  In our own 
Galaxy, OH masers have typical luminosities of $2\times10^{21}$\,Watts and 
sizes of 2 to 20 AU (\citealt{minier02}).
  At the other end of the scale are the so-called megamaser 
sources in active galaxies, sources a factor of $10^{7}$ brighter 
than ordinary Galactic masers, such as that discovered in Arp220 
(IC\,4553; \citealt{baan82}).  In between these extremes of the phenomenon are 
masers such as those detected previously in M82 (\citealt{weliachew84}).  
These have observed intensities of $\sim$10$^{25}$\,Watts and are sometimes 
referred to as ``kilomasers" (e.g. \citealt{hagiwara01,henkel90}).  Studies of kilomaser 
sources in individual galaxies, such as M82, provides a sample of masers which are all at an 
approximately equal distance and hence can be observed with the same linear resolution.

This paper describes an investigation into the main line OH maser population in the nearby 
starburst galaxy M82 using data from both the Very Large Array (VLA) and Multi-Element Radio 
Linked Interferometer Network (MERLIN).  The maser population is investigated and their 
positions and velocity distribution are compared to other known objects and the distribution 
of other gases.

\begin{table}
\centering
\begin{tabular}{|ccccc|}
\hline
Telescope	& Date of 	& Angular 	& Bandwidth	& Velocity	\\
used		& observation	& resolution	& 		& resolution	\\
\hline
MERLIN		& 1995 Nov.	& 0\farcs2	& 8\,MHz	& 23\,\kms	\\
MERLIN		& 1997 May	& 0\farcs2	& 2\,MHz	& 1.4\,\kms	\\
VLA		& 2002 Apr/May	& 1\farcs4	& 6.3\,MHz	& 18\,\kms	\\
\hline
\end{tabular}
\caption[Summary of observations of OH masers in 
M82]{\label{masers_obssum}Summary of the observations of OH masers in M82.}
\end{table}

\section{Masers in the M82 starburst}

M82 is one of the closest, and therefore best studied, starburst galaxies
and radio observations of this galaxy are numerous.  Extensive studies of
the distribution and dynamics of neutral hydrogen have been made (e.g.
\citealt{wills00, wills02}) and these have been compared with the 
molecular gas as traced by CO emission (\citealt{shen95}).  The angular 
resolution of the CO observations is, however, comparatively low.  
Observations of the transitions of the OH molecule can be made with the 
same instruments at similar resolution to the previous H{\sc i} studies 
and so provide a better comparison of atomic and molecular gas.

An OH maser was first detected in M82 using the Effelsberg telescope by 
\cite{nguyen76}.  Using the VLA, \cite{weliachew84} detected two main line 
maser regions in the galaxy and noted that, although both spots are an 
order of magnitude brighter than the most luminous Galactic maser, the 
emission from those in M82 may be coming from a number of smaller masers 
within each spot.  The resolution of these observations however was not 
sufficient to detect any substructure or extension in either masing 
region.

Subsequent observations have resulted in further detections of masers at 
other frequencies.  A total of six OH features were detected using the 
1612 and 1720\,MHz satellite lines by \cite{seaquist97}.  Four features 
were seen in emission at 1720\,MHz, two were seen in absorption at the 
same frequency, while only one feature was detected in emission at 
1612\,MHz at the same position as one of the 1720\,MHz absorption features.  
H$_2$O masers have also been detected in M82 at 22\,GHz 
(\citealt{baudry96}; \citealt{hagiwara05}).


Recently MERLIN and the VLA have been used to probe the OH absorption 
across the central starburst region in M82.  As well as deep absorption features, eleven 
main line OH masers have been detected to a limit of 3$\sigma$, nine of which are new 
detections.

Unless otherwise stated, all positions are given in J2000 
coordinates in the format aa\farcs{\rm aaa} bb\rasec{\rm bb} corresponding 
to 09$^{\rm h}$55$^{\rm m}$aa\rasec{\rm aaa} and +69\degr40'bb\farcs{\rm 
bb} respectively.  At the distance of M82 (3.2\,Mpc, \citealt{freedman94}), one arcsecond 
corresponds to a linear distance of 15.5\,parsecs.

\begin{figure*}
\centering
\begin{tabular}{cc}
\includegraphics[width=6cm]{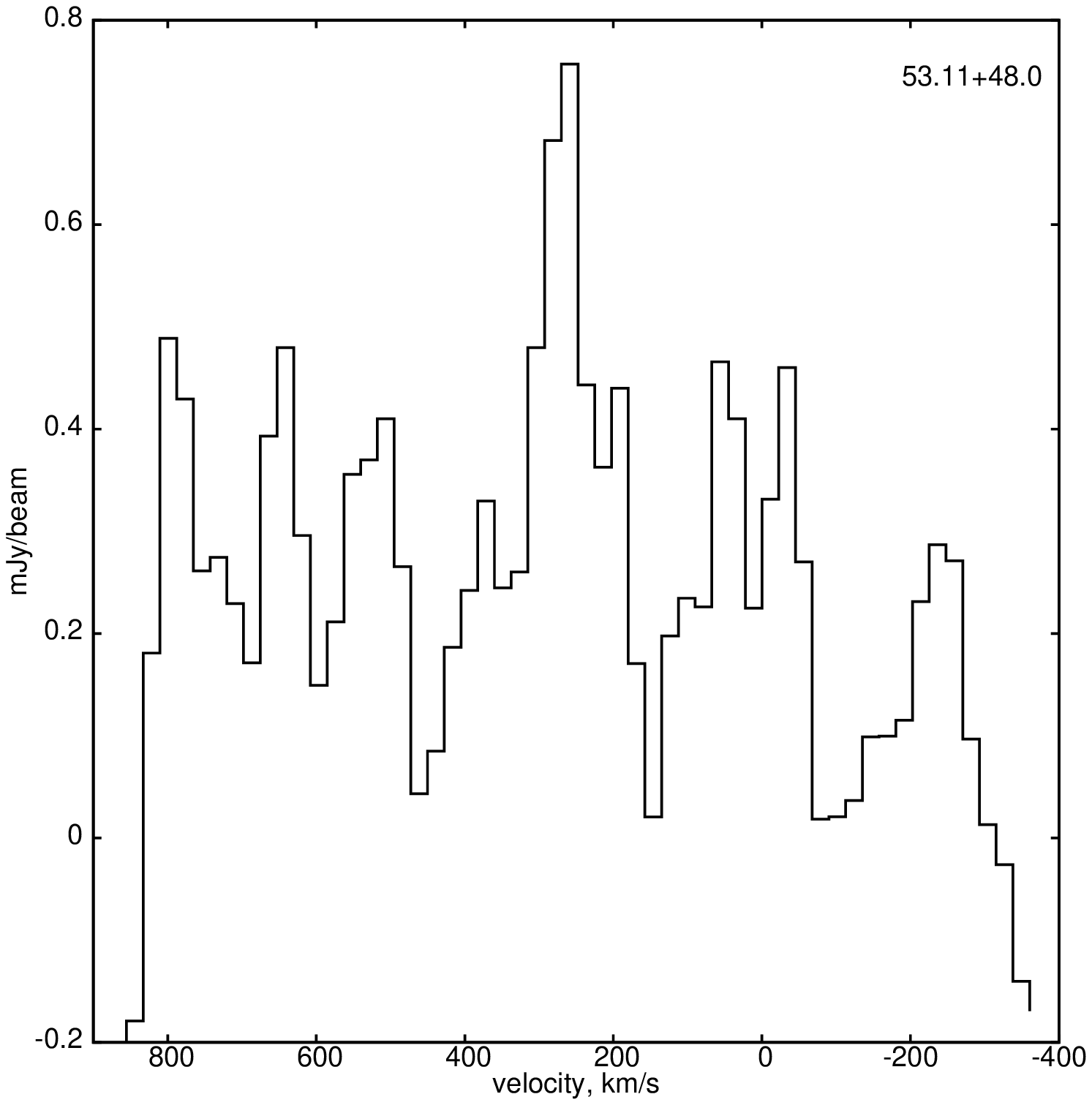}	 &
\includegraphics[width=6cm]{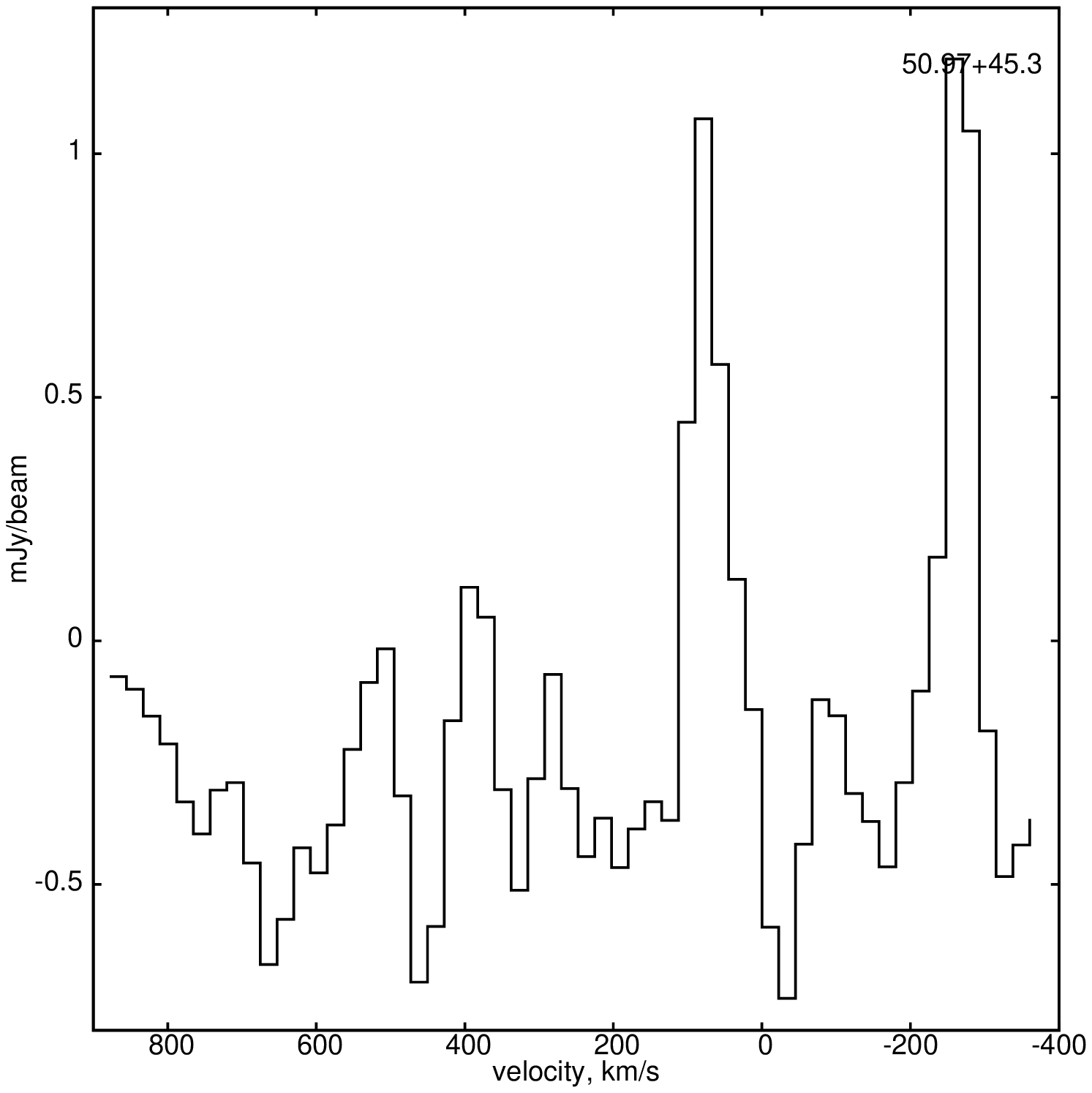}	\\
\includegraphics[width=6cm]{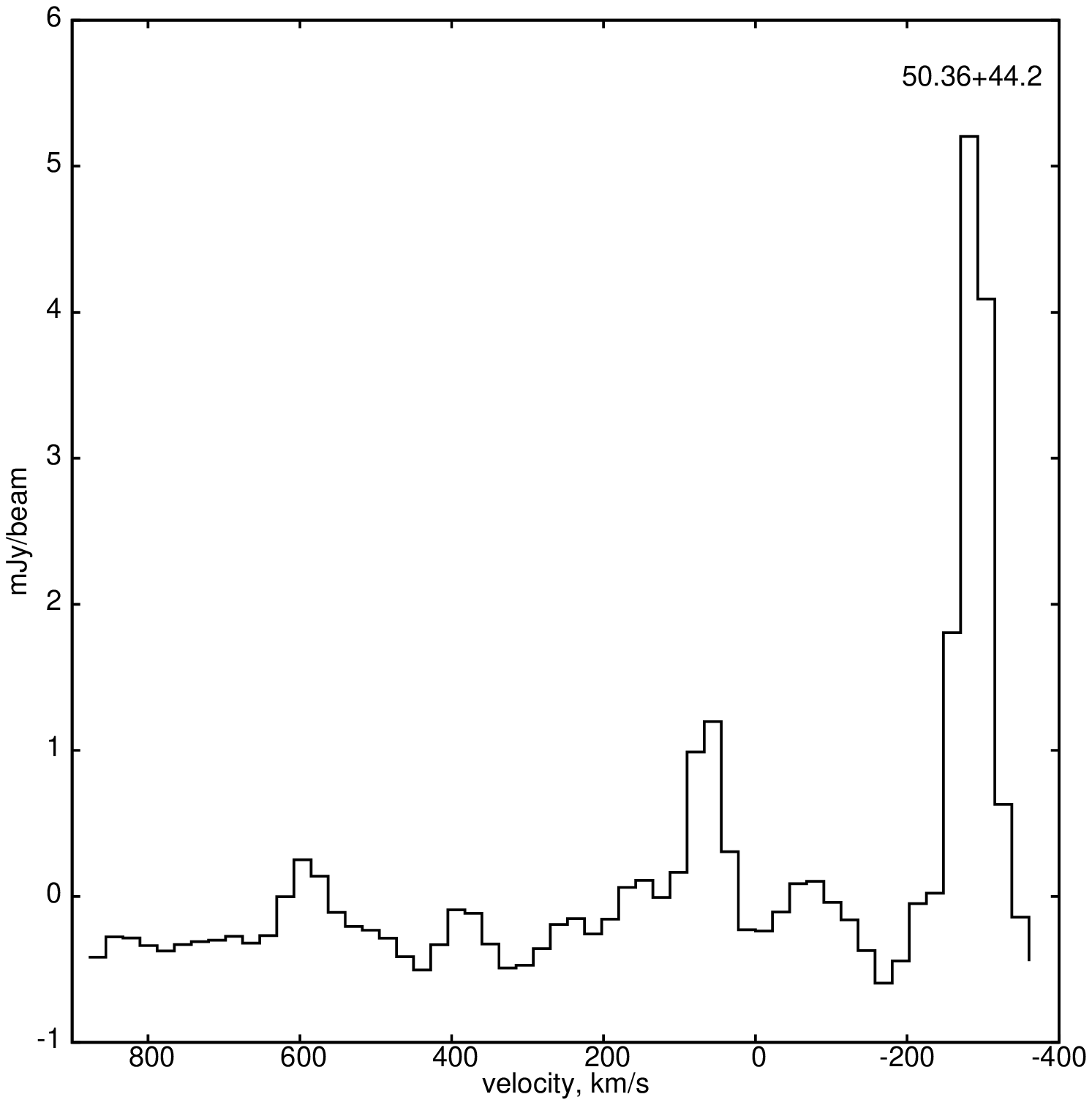}	 &
\includegraphics[width=6cm]{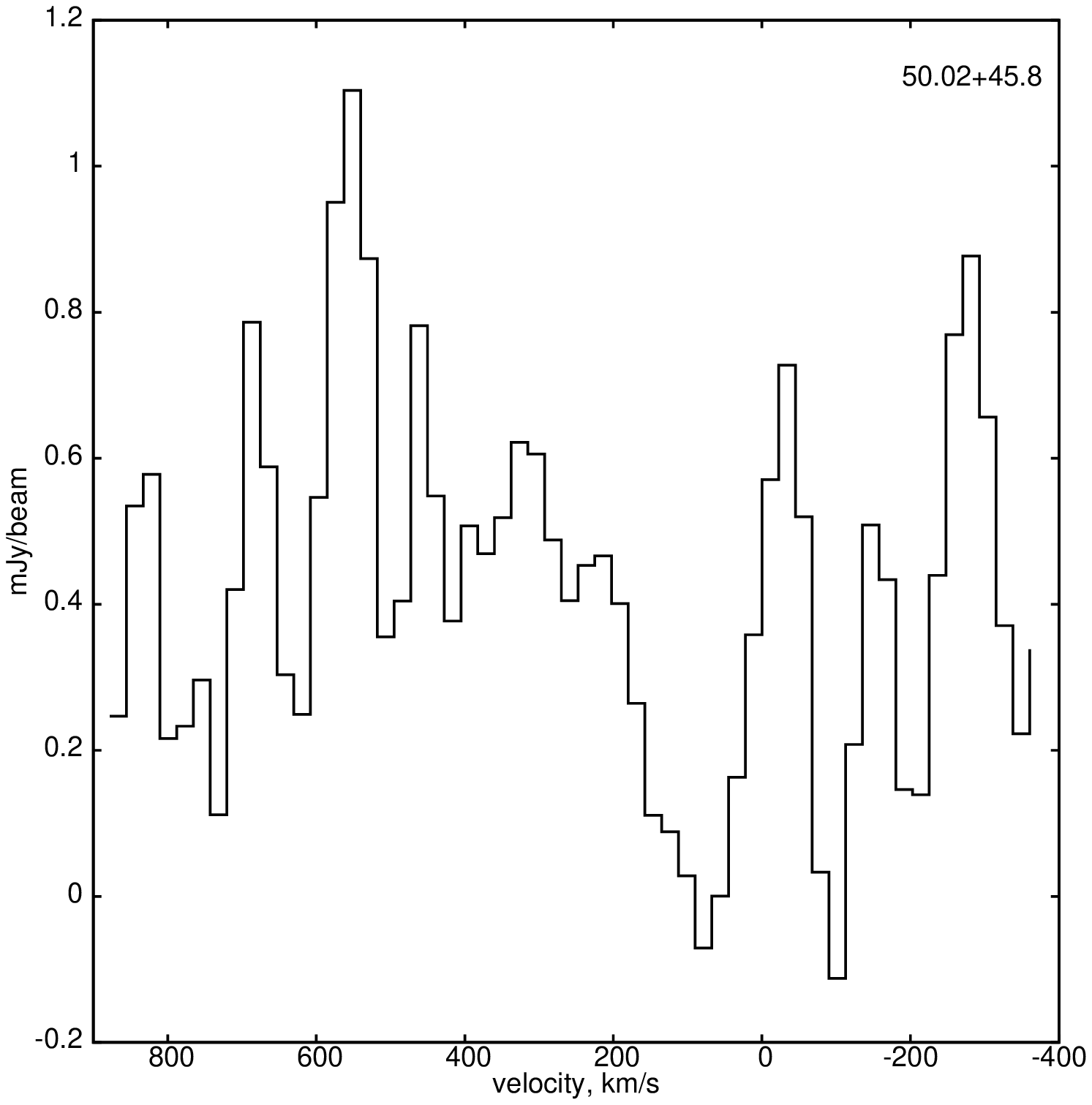}	\\
\includegraphics[width=6cm]{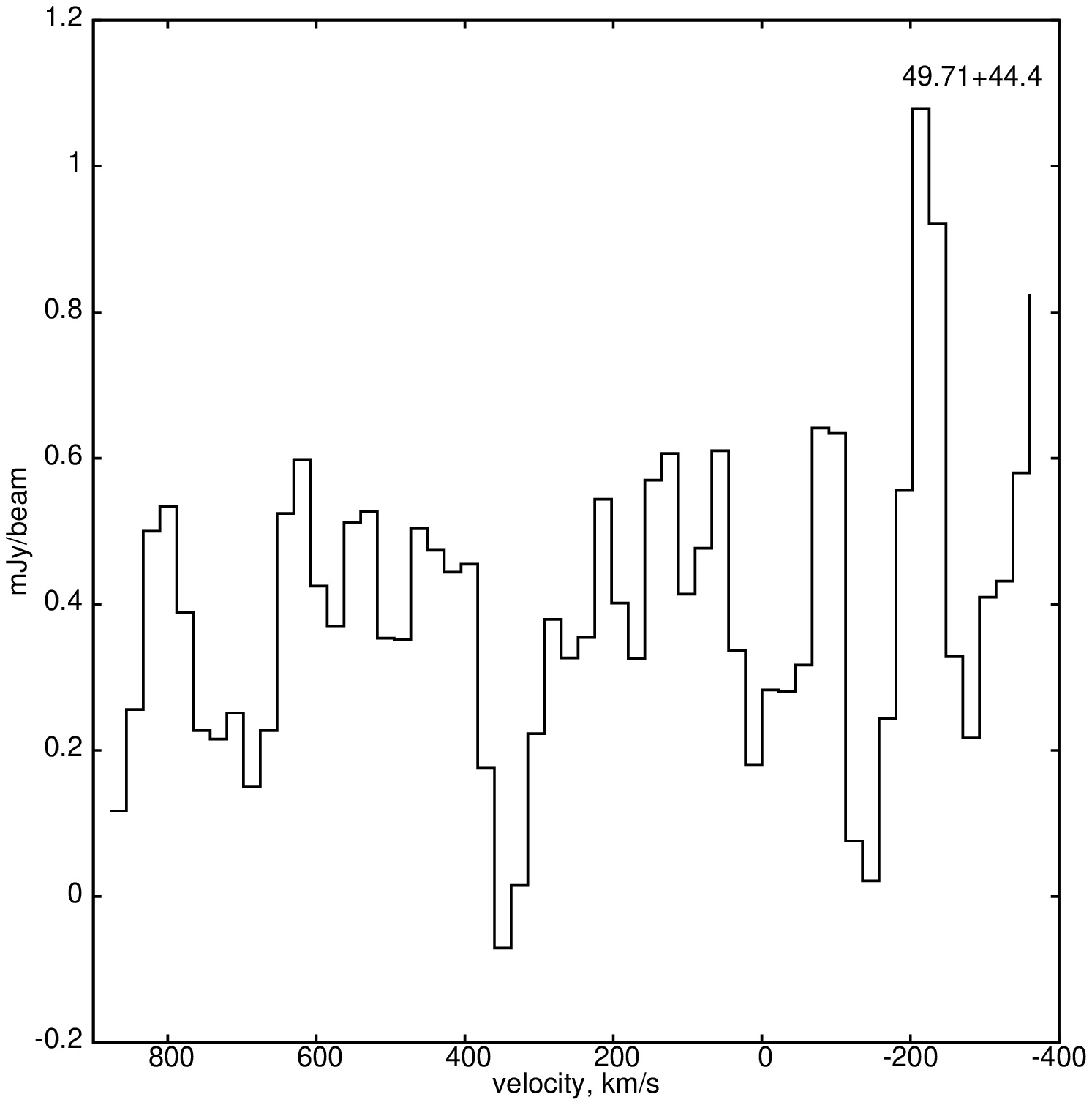}	 &
\includegraphics[width=6cm]{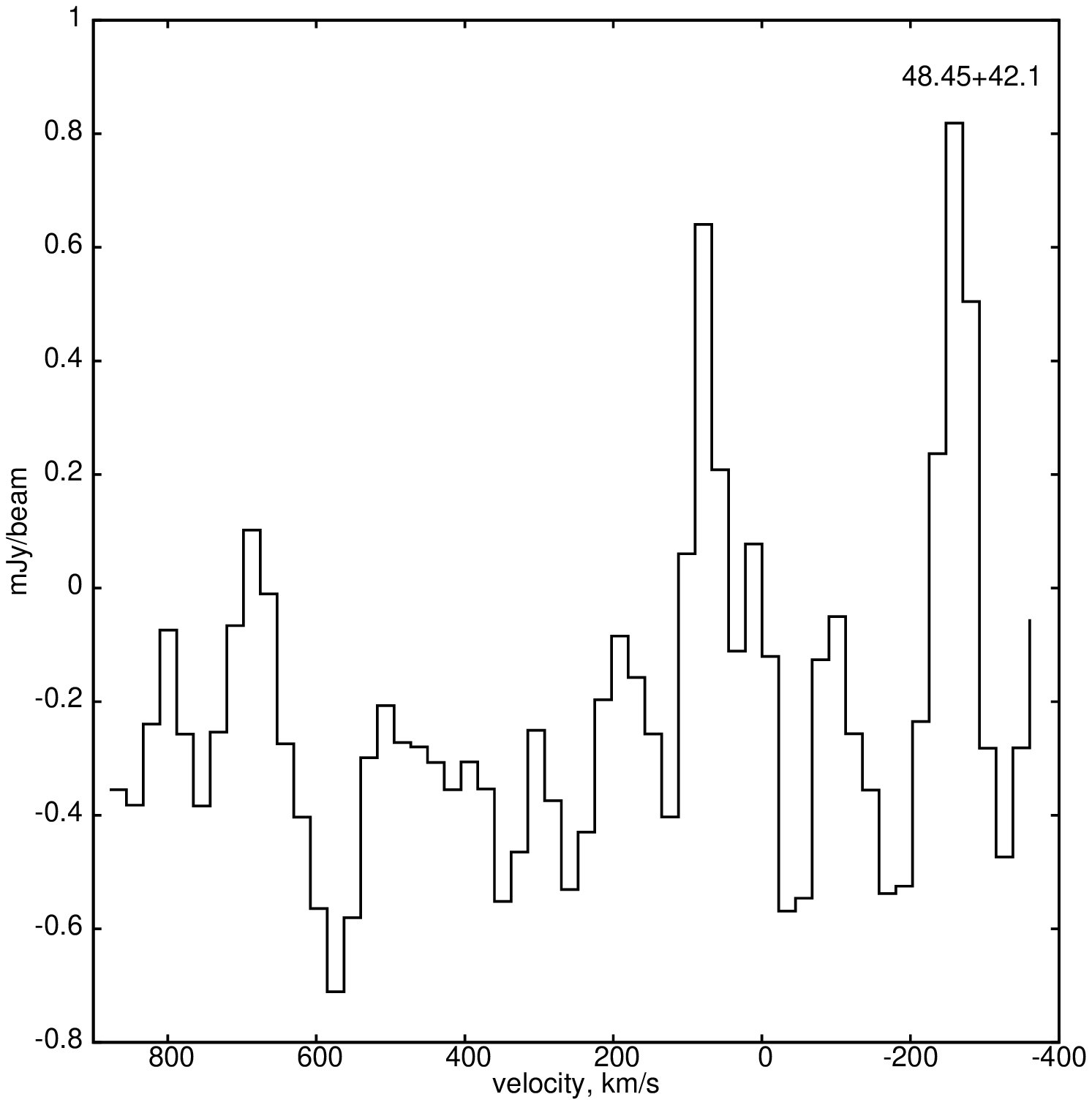}	\\
\end{tabular}
\caption[OH maser emission in the MERLIN 1995 observations]
{\label{fig_merlin95}Spectra of the maser features detected in the 1995 observations 
with MERLIN.  All are detected at $>$3$\sigma$ in at least one line.  The 
labels in the top right of each panel correspond to the IDs given in Table 
\ref{merlin95Table}.  The 3$\sigma$ noise level in each channel is 
2.59\,mJy/beam and the spectra shown here have been smoothed in frequency 
using a Gaussian of width two channels.  The $x$-axis velocity scale is calculated 
relative to the rest frequency of the 1665\,MHz line.}
\end{figure*}


\section{Observations}

Data from the MERLIN archives, as well as more recent VLA 
observations, have been used to investigate the maser population in M82.  
A summary of the observations discussed here is given in Table 
\ref{masers_obssum} and further details are given in the following Section.


\subsection[MERLIN archive]{MERLIN archive}

\subsubsection{1995 observations}	

MERLIN observations of OH in M82 were carried out 
on 26th November 1995 using a total bandwidth of 7.75\,MHz over 64 
channels giving a velocity resolution of $\sim$22\,\kms at a wavelength of 18\,cm.
Eight of the MERLIN telescopes were included in the array for these 
observations: Lovell, Mk 2, Tabley, Darnhall, Knockin, Defford, Cambridge 
and Wardle (the Wardle antenna has since been decommissioned).  This 
provided a maximum baseline of 217\,km and hence a resolution of 0\farcs2 
at a wavelength of 18\,cm.  Gaussian fitting of components using the AIPS task {\sc jmfit} resulted in a position 
uncertainty, relative to the phase calibrator, of $\pm$0\farcs02.

This was a preliminary observation intended to detect the two known 
masers, look for additional sources of OH emission, and also to probe 
absorption by OH in the central starburst in M82 to provide a comparison 
with that detected in CO emission (for example).  Due to the $u$-$v$ 
coverage of MERLIN, however, extended absorption features are resolved out 
and very little absorption was actually detected in these observations.  
Several emission features were detected, however.  Figure 
\ref{fig_merlin95} shows spectra of the masers detected in this 
observation.
Table \ref{merlin95Table}  lists the masers found in 1995 MERLIN data 
together with their positions and flux densities in this data-set.  The ID label 
for each maser, given in column one, is constructed from the VLA 2002 
position where known.  The exception to this is 50.02+45.8 whose ID is 
determined from the MERLIN positions listed in Table \ref{merlin95Table} 
as it was not detected in the 2002 data.  Note that the spectra are shown smoothed with a Gaussian of width two 
channels, while the peak flux densities listed in Table \ref{merlin95Table} are measured from the unsmoothed data.

\begin{table}
\begin{center}
\begin{tabular}{|c|cc|cc|c|}
\hline
ID	& RA	& Dec   & S$_{1665}$	& S$_{1667}$	& Ratio \\
	&	&	& (mJy)		& (mJy)		& \\
\hline
53.11+48.0 & 53\rasec094 & 47\farcs88 & 5.57& $<$2.6& $<$0.47\\  
50.97+45.3 & 50.924 & 45.37 & 8.35 & 7.83   & 0.94	\\  
50.35+44.1 & 50.350 & 44.13 & 7.01 & 29.9   & 4.27	\\  
50.02+45.8 & 50.017 & 45.83 & 4.30 & 5.09   & 1.18	\\  
49.71+44.4 & 49.689 & 44.17 & $<$2.6& 5.29  & $>$2.03	\\  
48.45+42.1 & 48.360 & 41.93 & 4.15 & 6.94   & 1.67	\\  
\hline
\end{tabular}
\caption[List of masers from the MERLIN 1995 
data-set]{\label{merlin95Table}Definite maser detections made with MERLIN 
in 1995.  The 3$\sigma$ noise per channel was 2.6\,mJy/beam.  The IDs 
are constructed from the J2000 positions measured in the VLA 2002 
data-set.  The flux densities are given in mJy and the listed ratios are 
for S$_{1667}$:S$_{1665}$.}
\end{center}
\end{table}

\subsubsection{1997 observations}

\begin{figure*}
\centering
\begin{tabular}{cc}
\includegraphics[width=6cm]{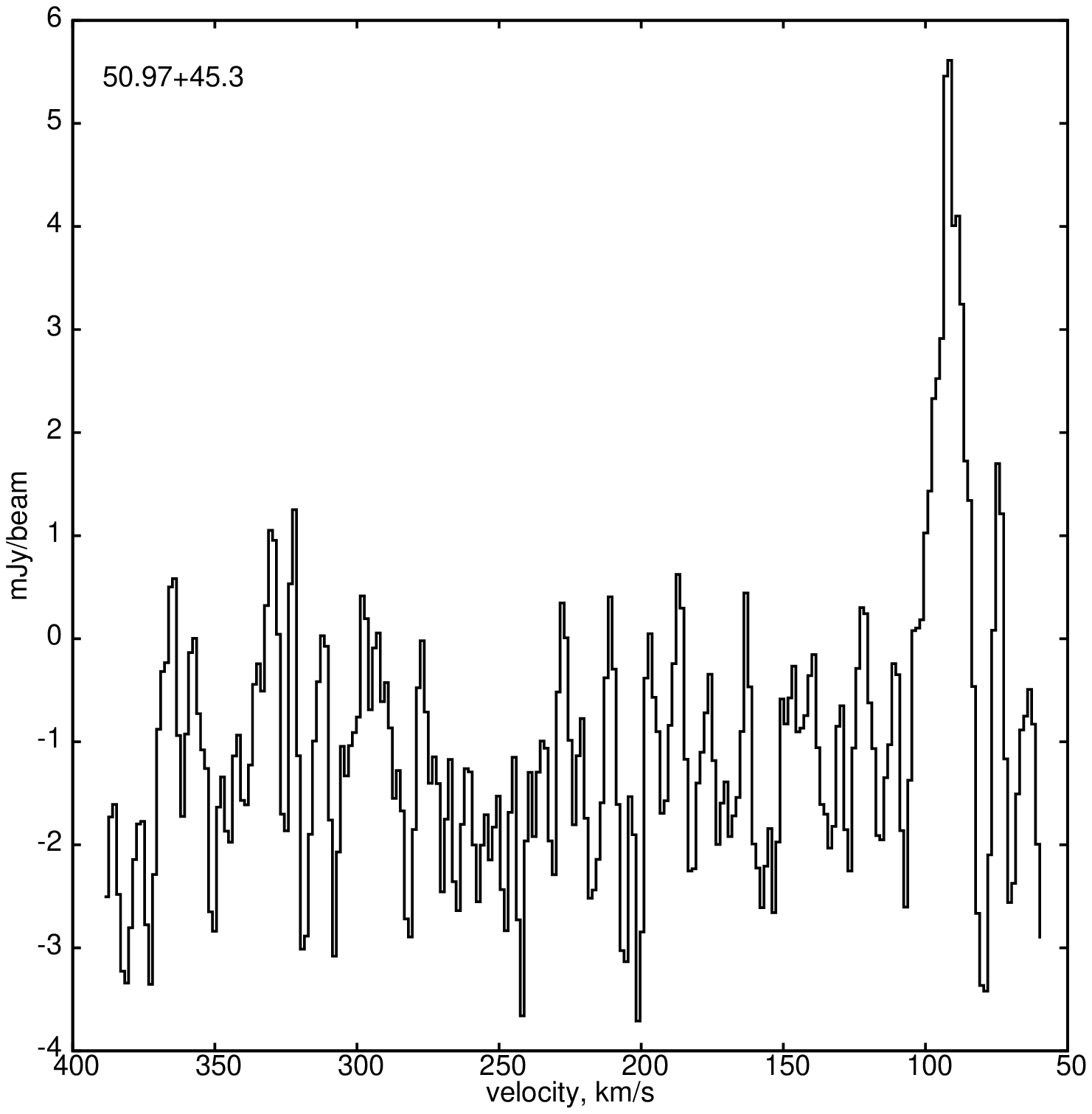}	&
\includegraphics[width=6cm]{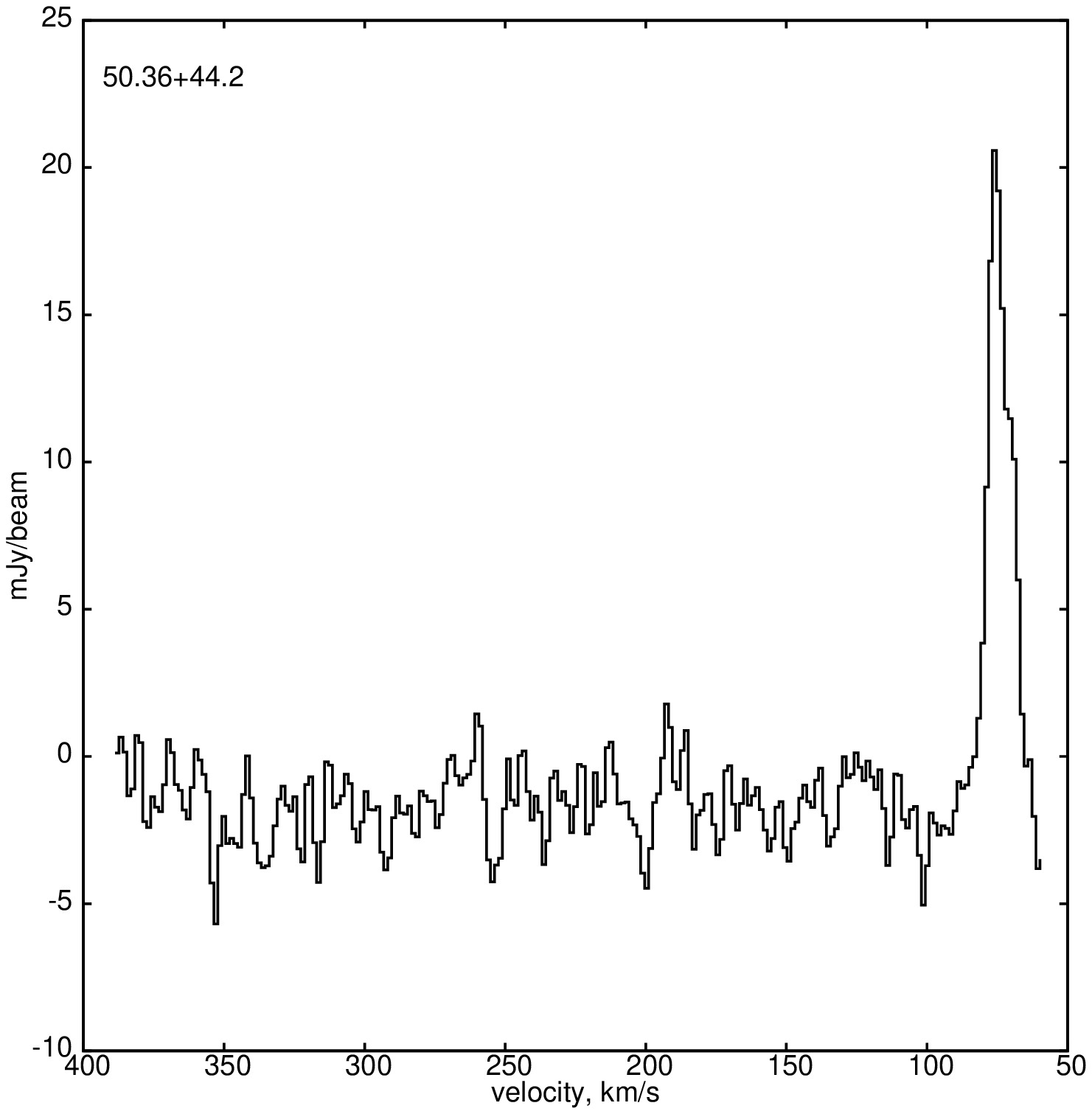}	\\
\end{tabular}
\caption[1667 maser emission in the MERLIN 1997 observations]
{\label{fig_1667merlin97}Spectra of the two maser features detected at 
1667\,MHz in the 1997 observations with MERLIN.  The 3$\sigma$ noise level 
in each channel was 18.2\,mJy/beam.  The data are shown smoothed in 
frequency using a Gaussian of width two channels.  The $x$-axis velocity scale is calculated relative to the rest 
frequency of the 1667\,MHz line.
}
\end{figure*}

These observations were carried out in May 1997 by switching between 1612, 
1665, 1667, and 1720\,MHz using seven telescopes of the MERLIN array: 
Defford, Cambridge, Knockin, Darnhall, Mk 2, Lovell and Tabley.  The 
correlator was configured to record 256 frequency channels 7.7\,kHz in 
width, giving a velocity resolution of 1.4\,\kms.  
Only the 1665 and 1667\,MHz observations are discussed here. 
Figure \ref{fig_1667merlin97} shows spectra of the two masers detected at 
1667\,MHz, illustrating the narrower frequency channel widths in this 
observation.  Gaussian fitting components in these date resulted in aposition uncertainty of $\pm$0\farcs03.

\begin{table}
\begin{center}
\begin{tabular}{|c|cc|cc|c|}
\hline
ID	& RA	& Dec   & S$_{1665}$	& S$_{1667}$	& Ratio		\\
	&	&	& (mJy)		& (mJy)		& \\
\hline
50.97+45.3 & 50\rasec924 & 45\farcs37 & $<$18 & 23.0 & $<$1.35 \\  
50.35+44.1 & 50\rasec384 & 44\farcs24 & $<$18 & 57.3 & $<$3.36 \\  
\hline
\end{tabular} 
\caption[List of masers from the MERLIN 1997 data-set]{\label{merlin97Table}
Definite maser detections made with MERLIN in 1997.  The flux densities 
are given in mJy and the ratio is S$_{1667}$:S$_{1665}$.  The 3$\sigma$ 
noise was 17\,mJy/beam in the 1665\,MHz observation and 18\,mJy/beam in 
the 1667\,MHz observation.}
\end{center}
\end{table}

As these observations were carried out using narrower frequency channels 
than the 1995 observations there is a reduction in sensitivity. 
The 3$\sigma$ noise level per channel in these observations was 
16.9\,mJy/beam at 1665\,MHz and 18.2\,mJy/beam at 1667\,MHz, the 
difference arising from the longer observation at 1665\,MHz.

As stated previously, the $u$-$v$ coverage of MERLIN is not ideal for 
detecting extended absorption and most of it was resolved out in these 
observations.  For detecting compact objects such as masers, however, 
MERLIN's resolution is an advantage as the observations are not greatly 
affected by deep absorption features.

The detections made in the 1667\,MHz data-set are listed in Table 
\ref{merlin97Table}.  No detections were made in the 1665\,MHz data-set.  
In most cases in the other data-sets examined, the masers are 
significantly weaker in this line compared to the 1667\,MHz line.


\subsection{VLA 2002}

Observations were carried out in April and May 2002 using the VLA in A 
configuration.  
At 1.6\,GHz this gives an angular resolution of $\sim1\farcs4$.  The data 
were taken in two observing runs, one of which was carried out in April, 
the other in May 2002.  The data were recorded in 64 channels over a 
bandwidth of 6.25\,MHz such that both main line maser frequencies were 
within the observing band over the extent of the velocity range within the 
disk of the galaxy.  This setup resulted in a velocity resolution of 
18\,\kms.  3C286 (1331+305) was used for flux calibration, 0319+415 was 
used for bandpass calibration, and the phase calibrator was 0954+745.  
Calibration of the data was performed using standard techniques for the VLA.
In order to isolate the line component, the radio continuum emission was
subtracted using a set of line free channels near the start and end of the 
cube where the channel frequency fell outside that corresponding to the 
velocity range for OH in M82.

Some of the maser features are bright enough to cause ringing in the 2002 VLA data.
This effect can be largely removed by smoothing the data, although this 
results in adjacent channels no longer being independent.  This was 
carried out using the AIPS task {\sc xsmth} to apply a Gaussian function 
with a width of two channels.  This also has the effect of reducing the 
measured peak flux since the emission is effectively smeared over several 
channels, hence the peak fluxes listed in Table \ref{vla02Table} were 
measured in the unsmoothed data.

The OH absorption across the central starburst in M82 is well displayed in 
these observations, but the spectral resolution of the data-set is not 
ideal for maser studies.  Despite this, several maser features were 
detected in emission in these data.  Table \ref{vla02Table} lists the 
positions of the maser features in this data-set which are $>$3$\sigma$, 
while Figure \ref{fig_masers+ohabs} shows a plot of the maser positions on 
top of a contour plot of the OH absorption across the M82 starburst.
Gaussian fitting resulted in a position uncertainty of $\pm$0\farcs2.
Figure \ref{fig_2002spectra} shows spectra of each maser feature listed in 
Table \ref{vla02Table}.

\afterpage{\clearpage\begin{figure*}
\begin{center}
        \parbox{2.4in}{\psfig{figure=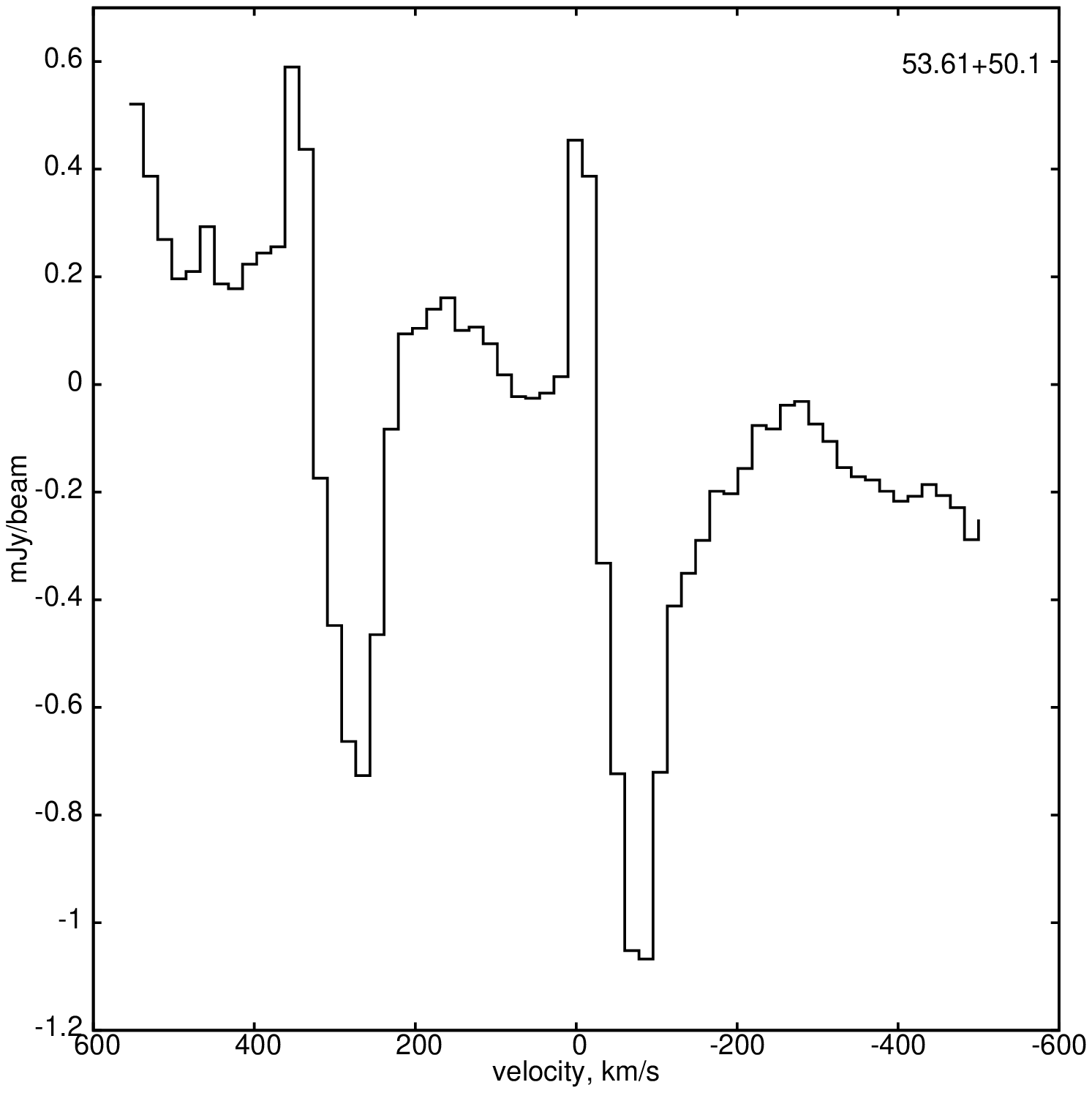, width=6cm}}
        \parbox{2.4in}{\psfig{figure=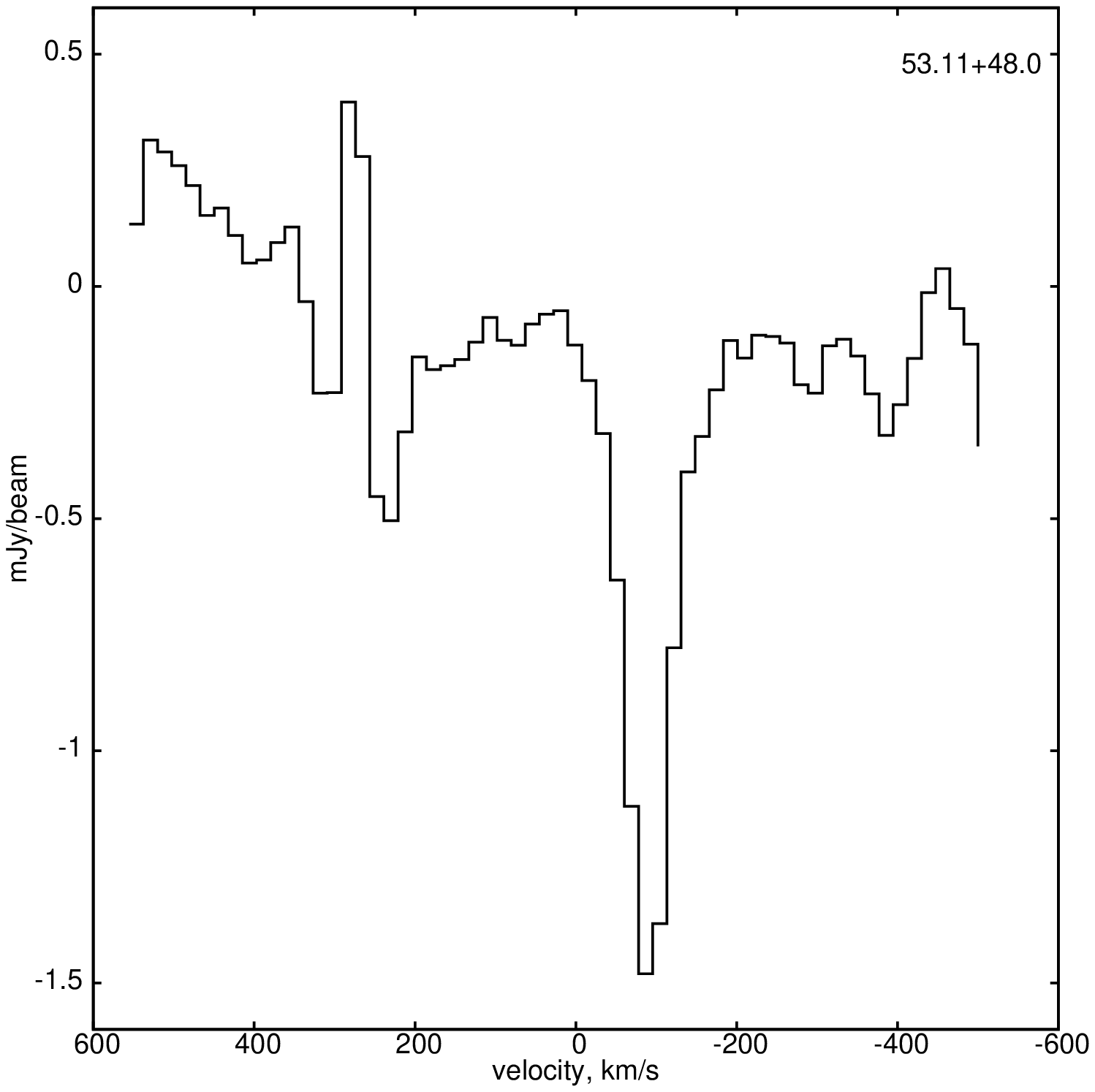, width=6cm}}
        \parbox{2.4in}{\psfig{figure=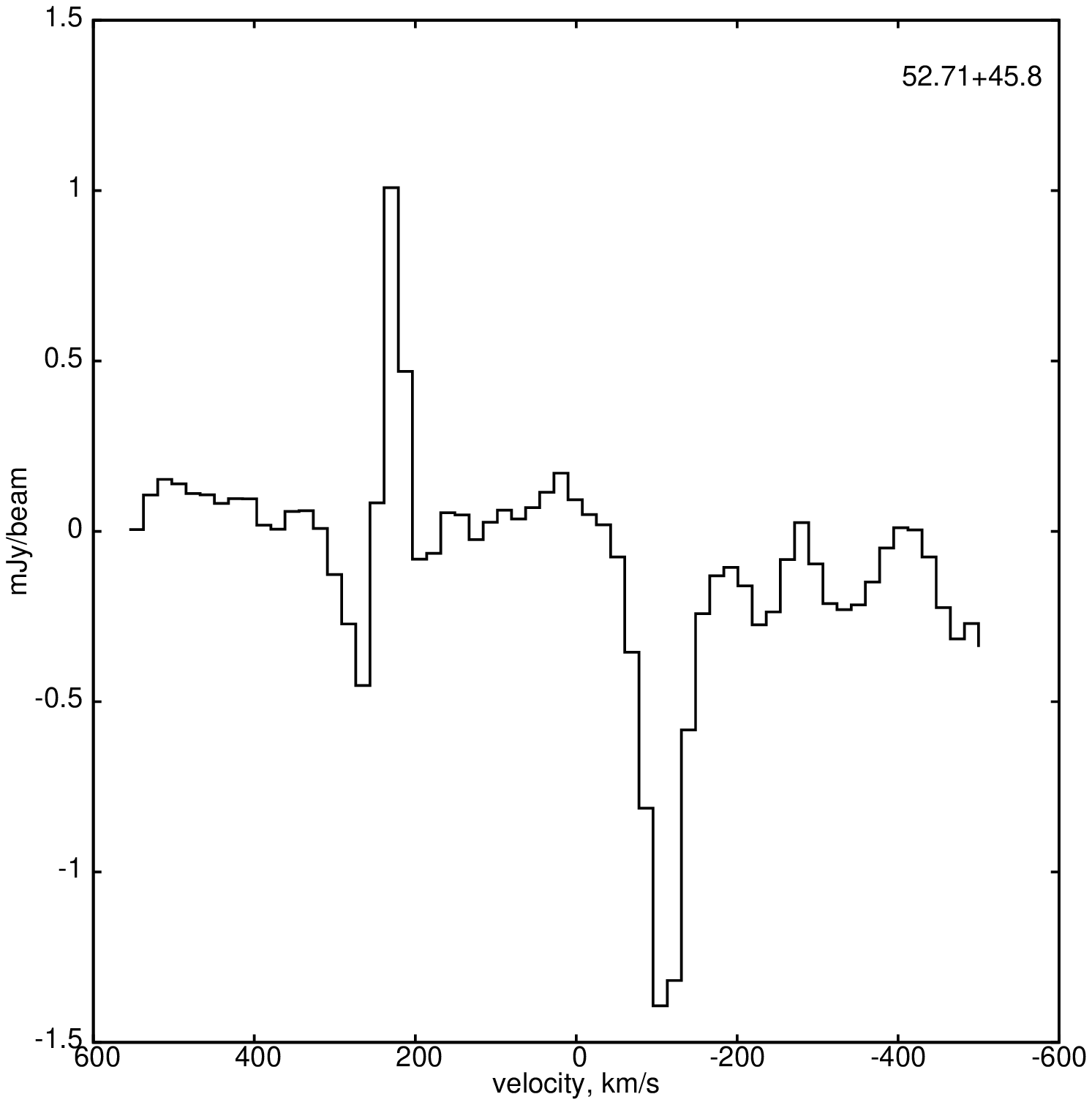, width=6cm}}
        \parbox{2.4in}{\psfig{figure=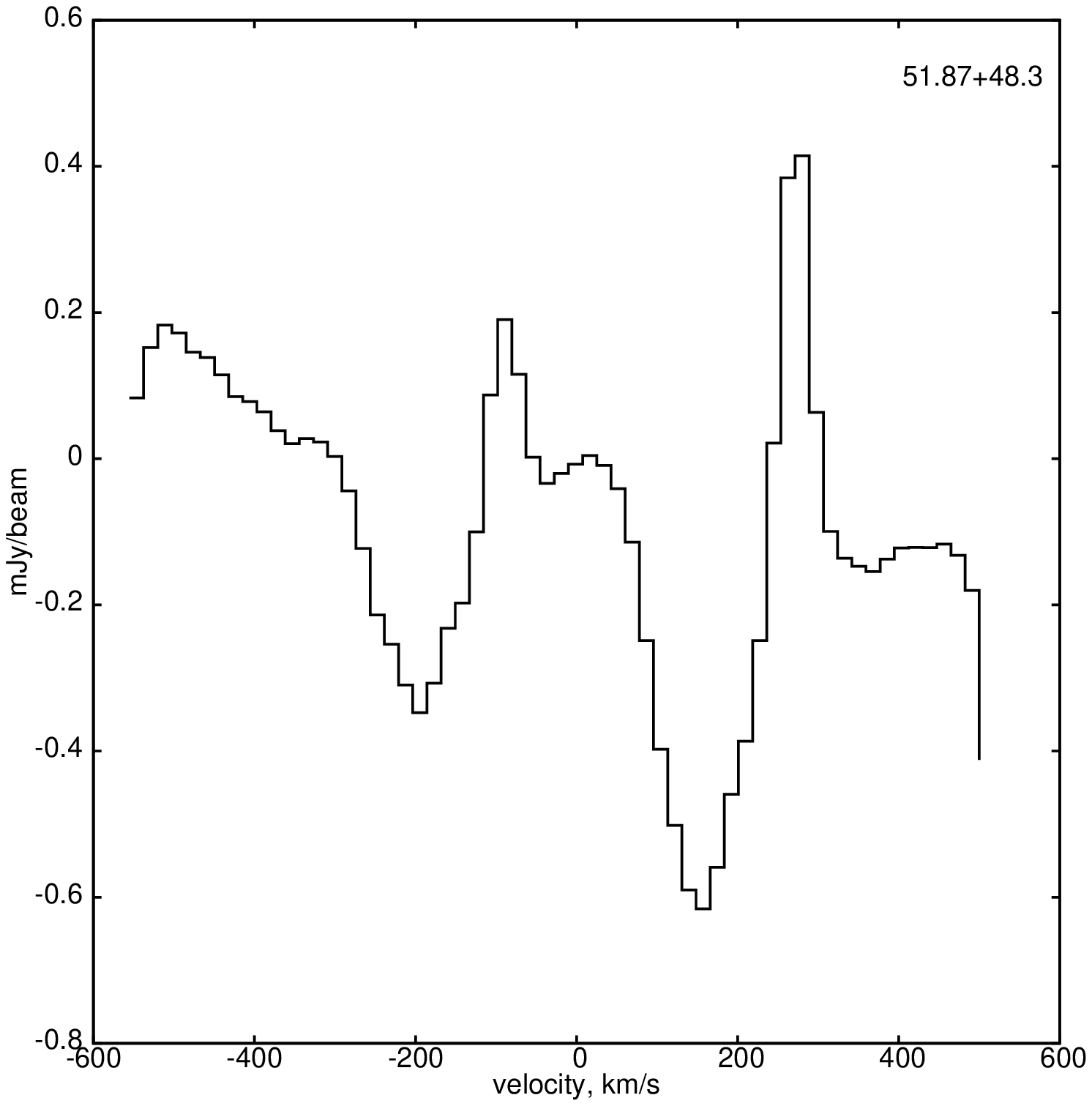, width=6cm}}
        \parbox{2.4in}{\psfig{figure=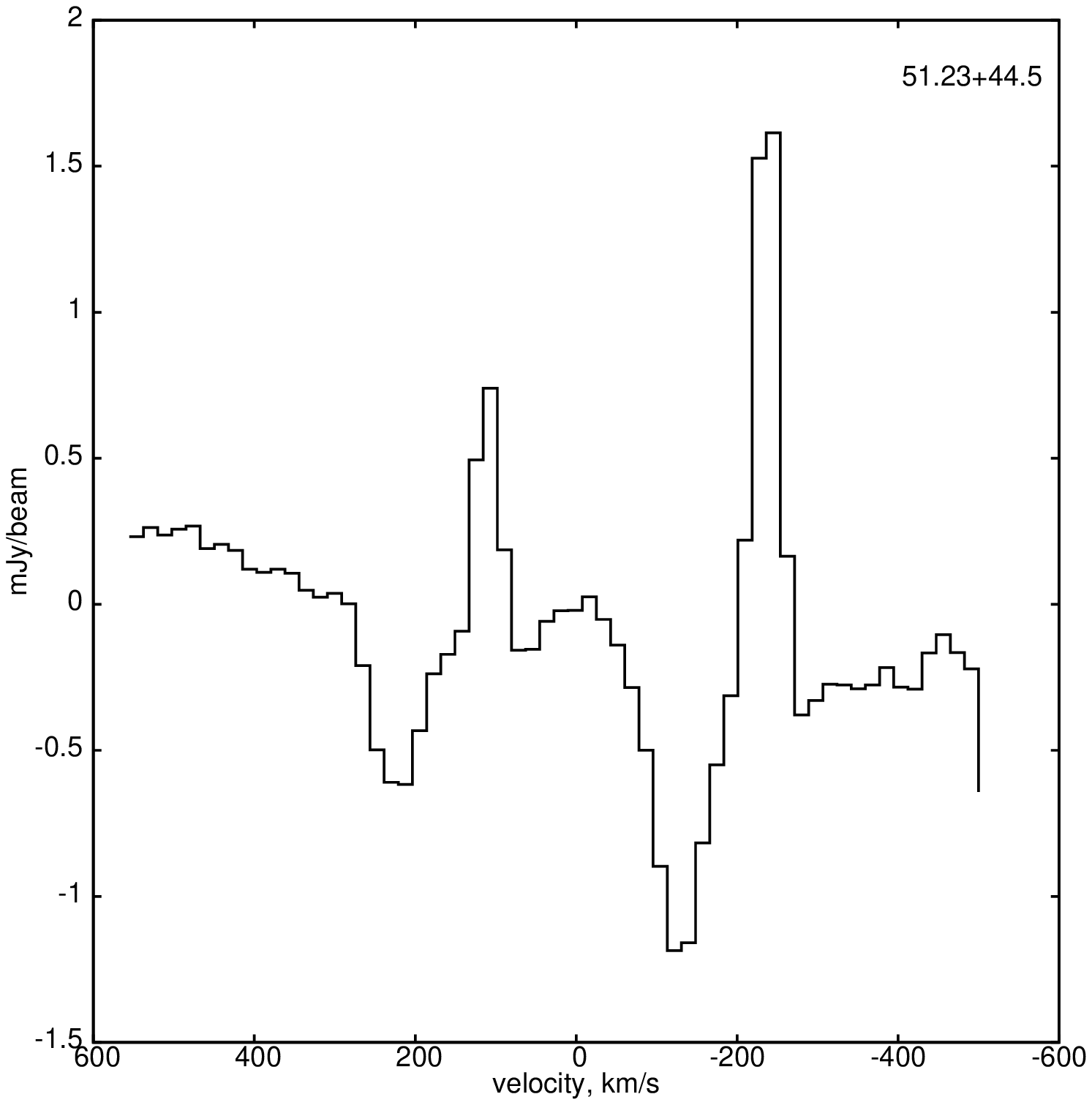, width=6cm}}
        \parbox{2.4in}{\psfig{figure=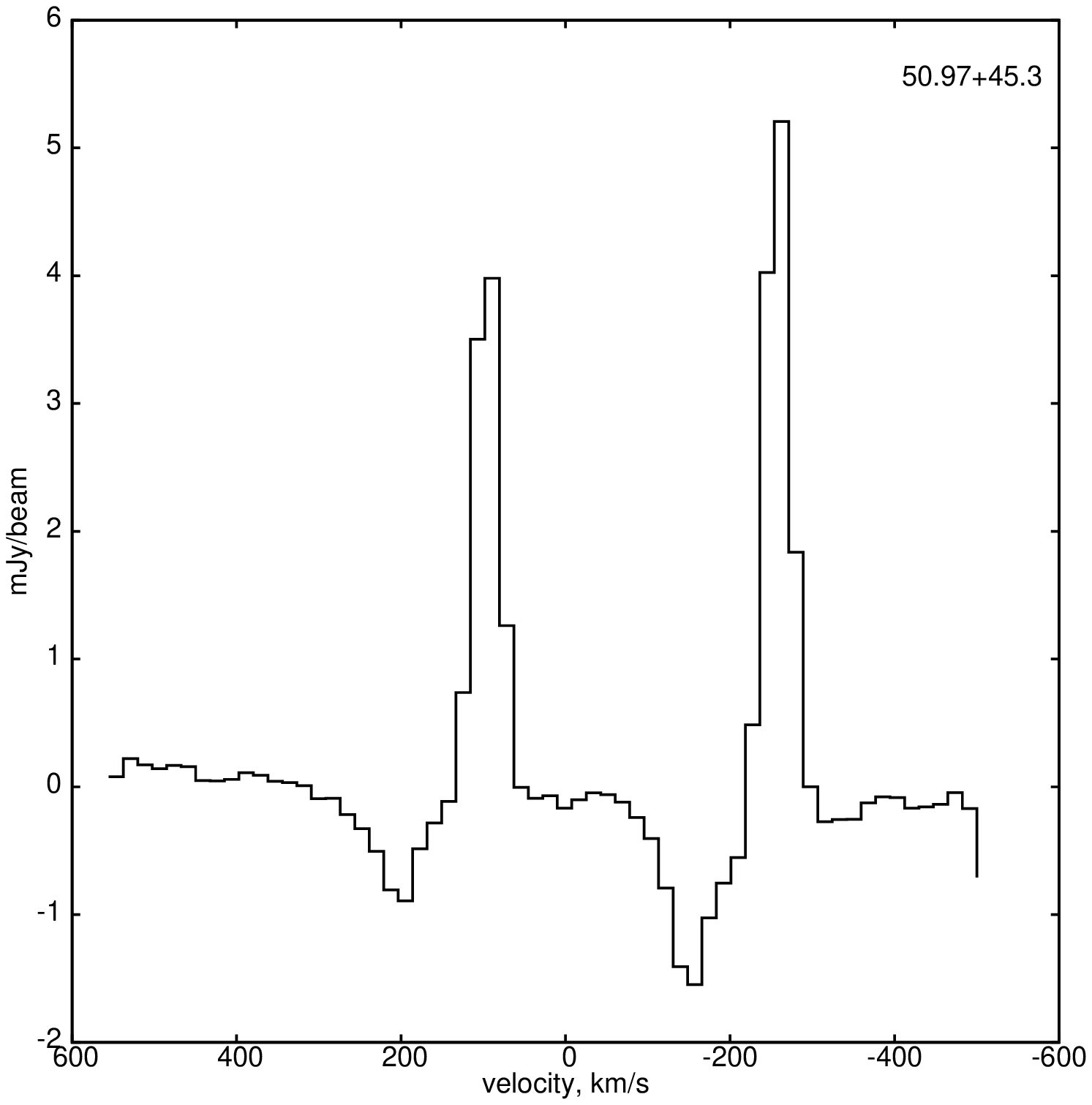, width=6cm}}
\end{center}
\caption[Spectra of masers from VLA 2002
observations]{\label{fig_2002spectra}Spectra of each maser listed in Table
\protect\ref{vla02Table}, arranged in order of decreasing R.A.  Each maser 
is seen as two or more adjacent channels with a flux greater than three   
times the rms noise level.  The data are smoothed with a Gaussian of width
2 channels.  The $x$-axis velocity scale is calculated relative to the rest frequency of the 1665\,MHz line.  
Continued on next page.}
\end{figure*}
\begin{figure*}
\begin{center}
        \parbox{2.4in}{\psfig{figure=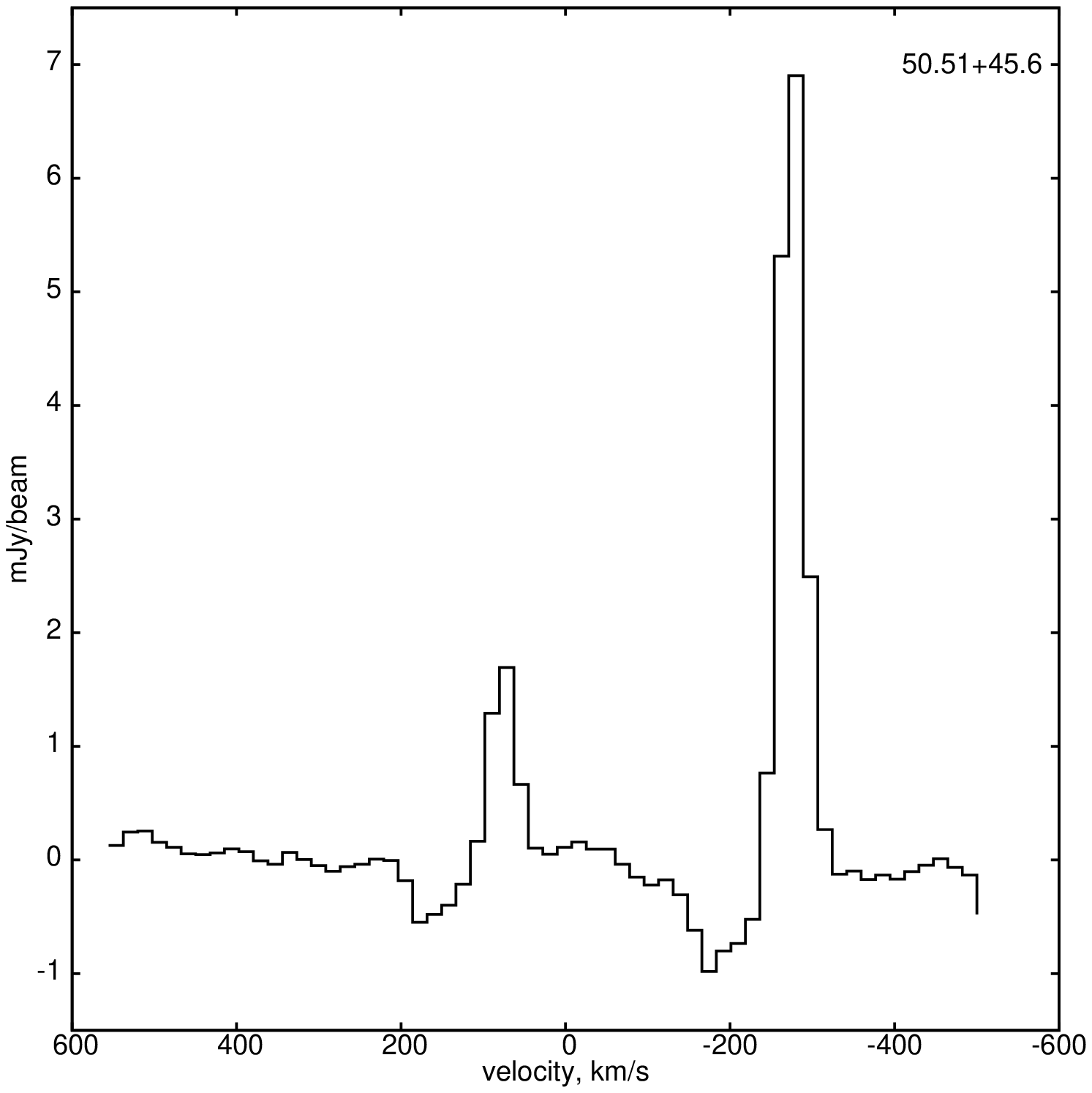, width=6cm}}   
        \parbox{2.4in}{\psfig{figure=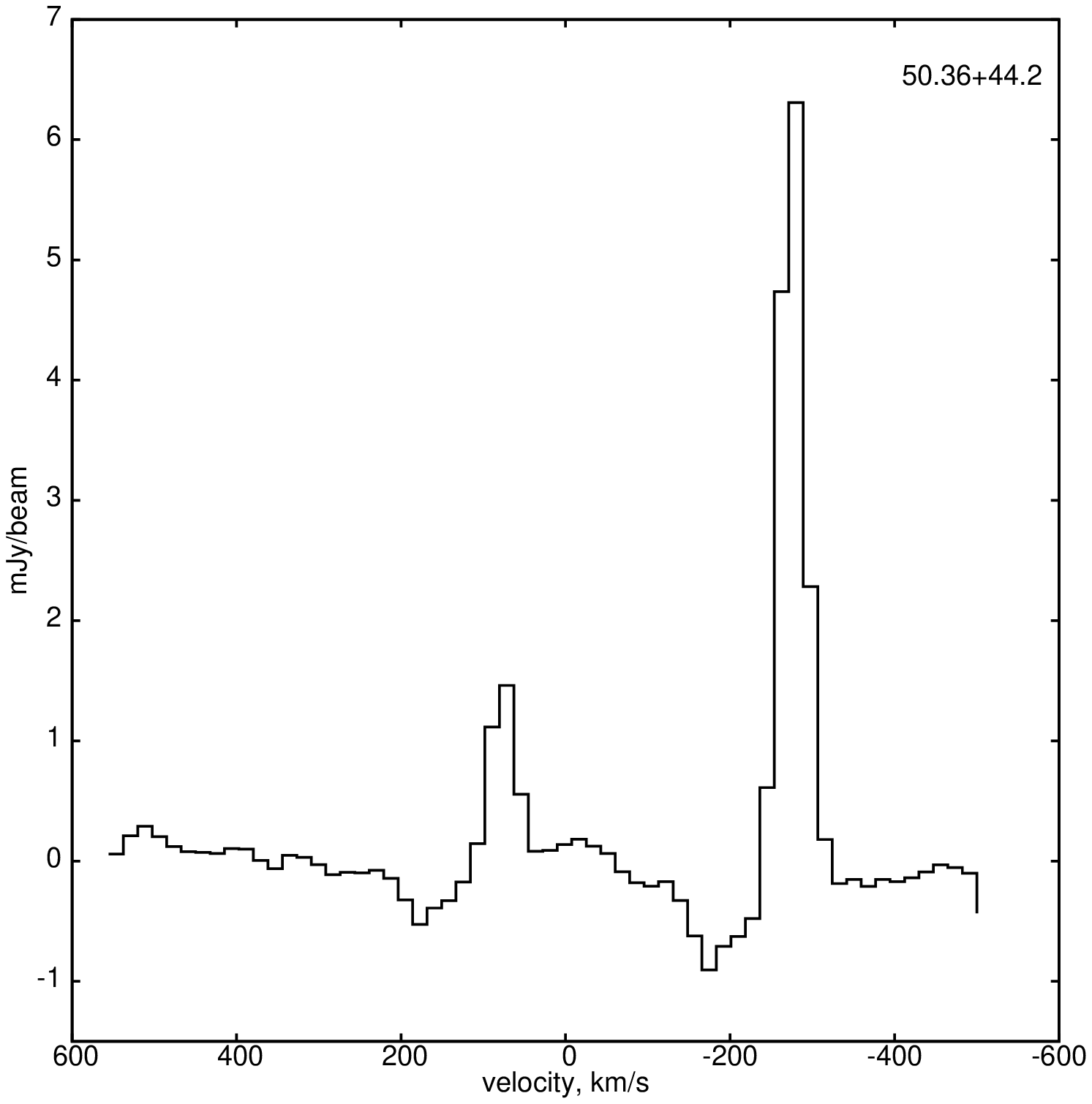, width=6cm}}   
        \parbox{2.4in}{\psfig{figure=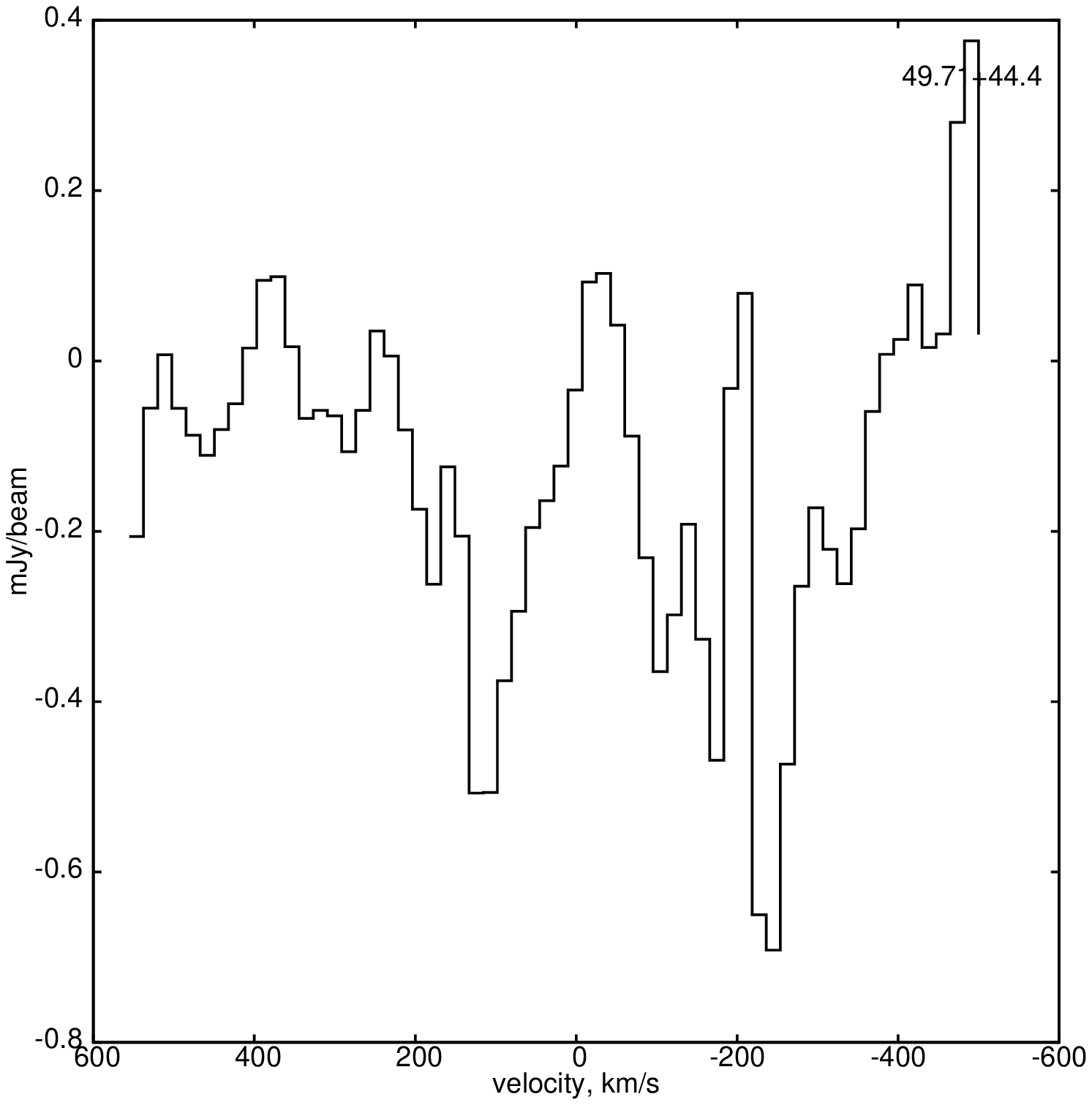,width=6cm}}  
        \parbox{2.4in}{\psfig{figure=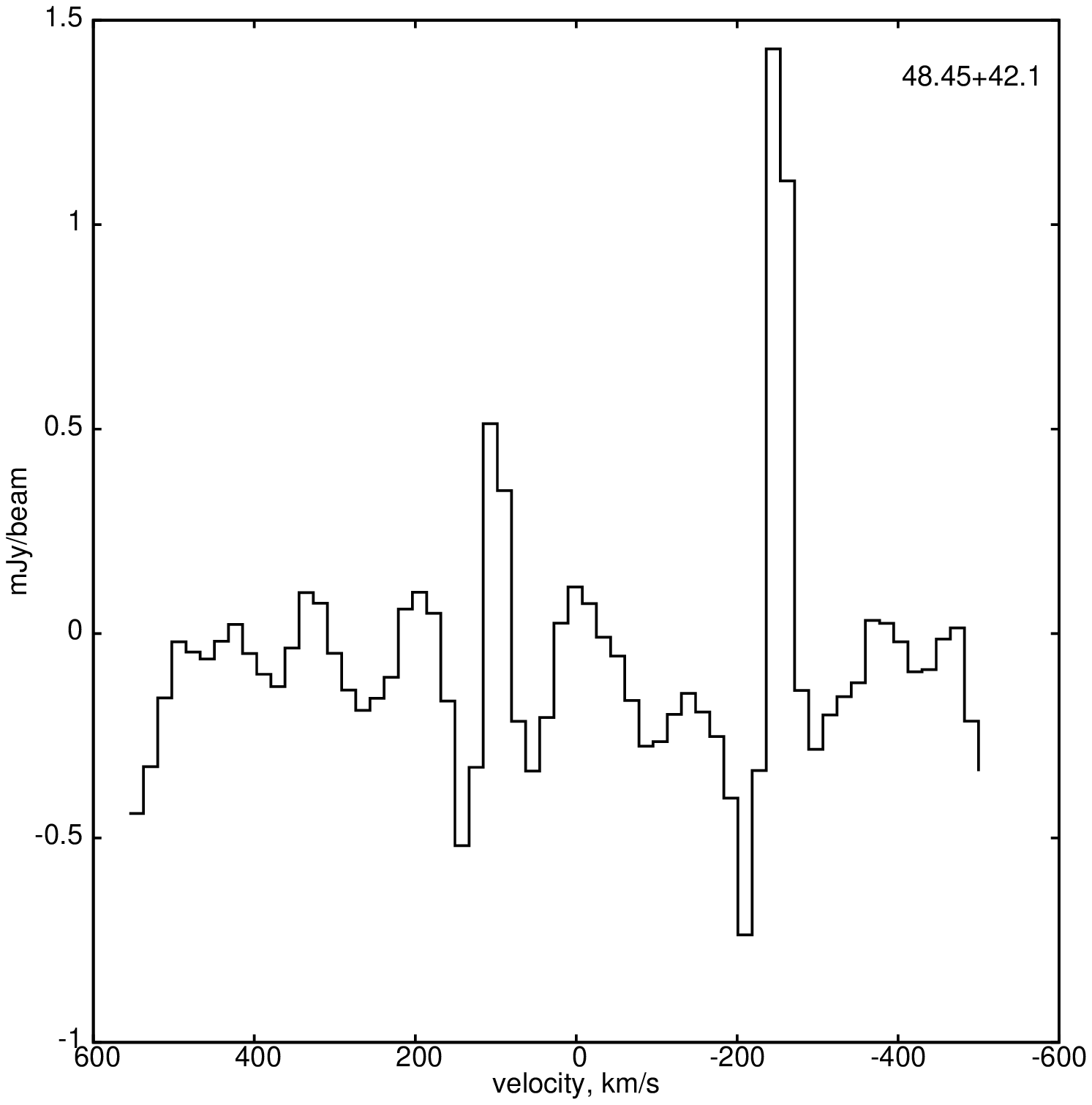, width=6cm}}\\
Figure \ref{fig_2002spectra}: continued.
\end{center}
\end{figure*}
}

The criteria used for classifying a signal as a maser was a detection of 
at least 3$\sigma$ in two or more consecutive channels.  This 
excludes some features which only appear in a single channel.  Due to the 
wide channels of this observation it is possible that some masers are 
missed altogether.  This could be because the emission is narrow but 
happens to fall at a channel edge, becoming spread over a couple of 
channels and hence too close to the noise level to be a firm detection.

Another major factor affecting the detection of a maser signal in these
data is the depth of the absorption.  As can be seen in Figure 
\ref{fig_2002spectra}, some of the masers are coincident with deep 
absorption features and are only detected because they are very bright.  
Weaker masers whose brightness is less than the magnitude of the depth of 
the absorption at the same position are therefore difficult to detect in 
these observations.

\begin{figure*}
\begin{tabular}{c}
\psfig{figure=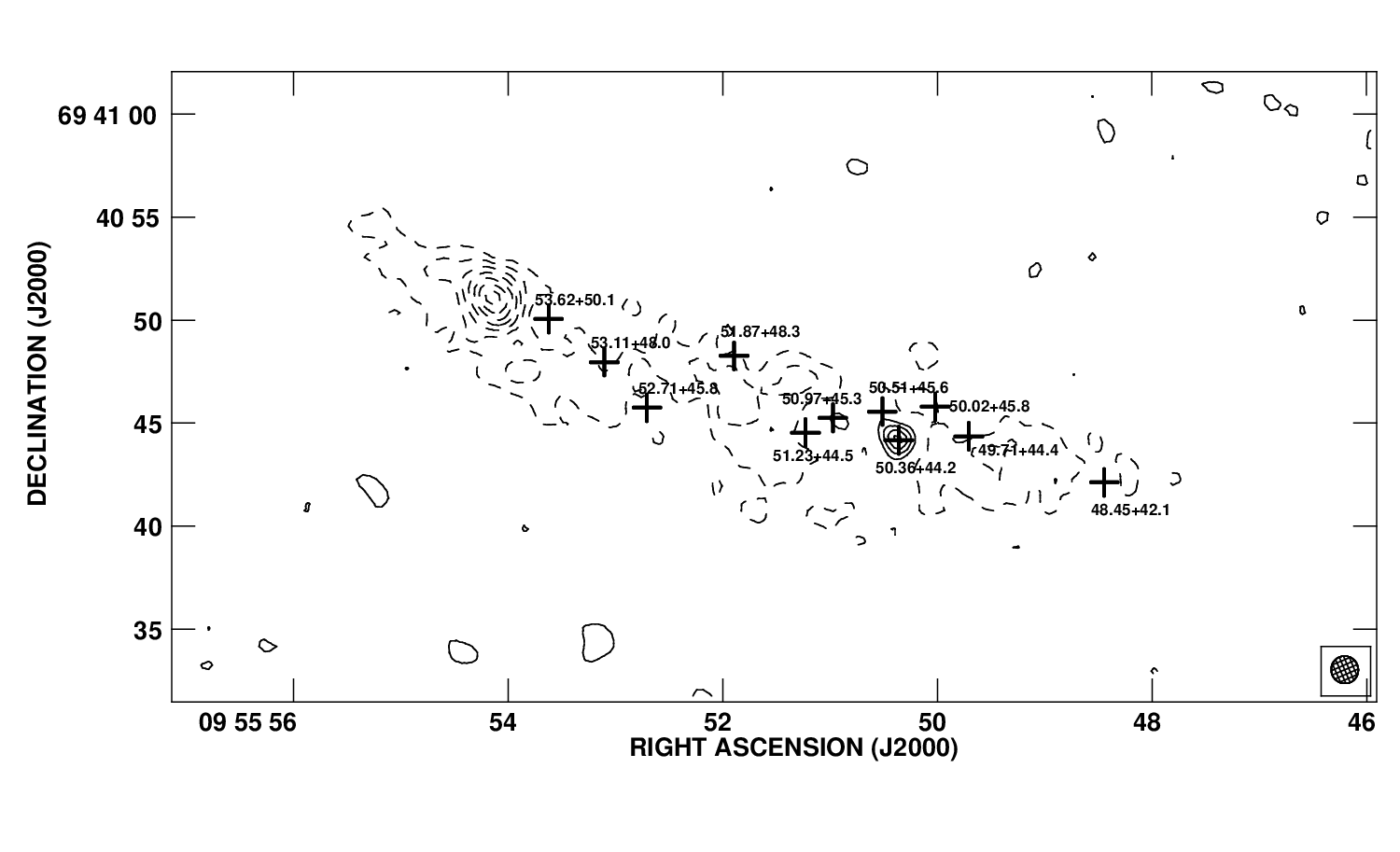,width=10cm,angle=270}
\end{tabular}
\caption[Masers overlaid on OH absorption in the M82 
starburst]{\label{fig_masers+ohabs}The location of the masers listed 
in Table \ref{vla02Table} superimposed on a contour plot of the OH 
absorption in the 2002 VLA data cube which has been continuum subtracted 
and summed in frequency using the AIPS task {\sc sqash}.
Contours are from -10 to +10 (in steps of 1) $\times$ 
0.2\,mJy\,beam$^{-1}$.  Solid contours are positive, dashed contours are 
negative.  The diameter of each symbol is that of the restoring beam, 1\farcs3.}
\end{figure*}

\begin{table*}
\begin{center}
\begin{tabular}{|c|cc|cc|c|}
\hline
ID	& R.A.	& Dec.   & S$_{1665}$	& S$_{1667}$	& Ratio (1667:1665)\\
	& (J2000)& (J2000)& (mJy)		& (mJy)		& \\
\hline 
53.62+50.1 & 53\rasec647 & 50\farcs09 & 2.44\,mJy & 2.49\,mJy	& 1.02	 \\  
53.11+48.0 & 53.105 & 47.97 & 1.86	& $<$0.77		& $<$0.41\\  
52.71+45.8 & 52.707 & 45.75 & 5.02	& $<$0.77		& $<$0.15\\  
51.87+48.3 & 51.868 & 48.31 & 1.71	& $<$0.77		& $<$0.45\\  
51.23+44.5 & 51.230 & 44.54 & 4.85	& 5.93 		& 1.22	\\ 
50.97+45.3 & 50.974 & 45.27 & 13.0	& 18.5		& 1.42	\\ 
50.51+45.6 & 50.513 & 45.57 & 1.85	& 1.28		& 0.692	\\ 
50.36+44.2 & 50.360 & 44.18 & 10.7	& 50.5		& 4.72	\\ 
49.71+44.4 & 49.707 & 44.37 & $<$0.77	& 2.34		& $>$3.04\\  
48.45+42.1 & 48.445 & 42.13 & 2.64	& 5.97		& 2.26	\\ 
\hline
\end{tabular} 
\caption[List of masers from the VLA 2002 data-set]{\label{vla02Table}
Definite maser detections identified from the VLA 2002 data.  Limits on 
the measured flux densities are given as 3$\sigma$.  The maser IDs here 
are constructed from the positions measured in this data-set and are used 
throughout this paper.
}
\end{center}
\end{table*}


\section{Notes on individual sources}

A brief description is given here for each source.  A more detailed 
discussion of the properties of the maser population is given in the 
following section.  {\bf Note} that, while the masers are identified here 
according to their J2000 positions, the continuum features to which they 
are compared (from \citealt{mcdonald02}) are identified using the B1950 
coordinate convention of \cite{kronberg75}.

\noindent{\bf 53.62+50.1} is coincident with an H{\sc ii} region 
(44.93+63.9) and was detected in both main lines in all but the 1997 
data-set.

\noindent{\bf 53.11+48.0} is coincident with a supernova remnant 
(44.43+61.9) and, while it was seen at 1665\,MHz in all but the 1997 
data-set, it was not detected at 1667\,MHz in any observation.

\noindent{\bf 52.71+45.8} is coincident with both a supernova remnant 
(44.01+59.6) and a satellite line maser (feature 4 from 
\citealt{seaquist97}).  Like 53.11+48.0, this maser is not detected at 
1667\,MHz in any observation.  It is detected at 1665\,MHz in the 
2002 data-set.

\noindent{\bf 51.87+48.3} is apparently not coincident with a known 
continuum feature or either of the other types of maser shown in Figure 
\ref{figMasers+}.  This source was only detected in the 2002 data and was 
one of the weaker detections.

\noindent{\bf 51.23+44.5} is located within an arcsecond of an H{\sc ii} 
region (42.48+58.4), and close to two water maser features (S1 and P3 
from \citealt{baudry96}) and two further H{\sc ii} regions.  Despite not 
being detected in either the 1995 or 1997 observations, it is seen in both 
lines in 2002.

\noindent{\bf 50.97+45.3} appears coincident with both an H{\sc ii} region 
(42.21+59.2) and a water maser feature (S2 from \citealt{baudry96}).  In 
the 1997 and 2002 observations this source is brightest at 
1667\,MHz, however in 1995 it was brighter at 1665\,MHz.

\noindent{\bf 50.51+45.6} is not clearly associated with another feature 
in M82 and is brighter at 1665\,MHz than 1667\,MHz.

\noindent{\bf 50.36+44.2} is closely associated with an H{\sc ii} region 
(41.64+57.9) and within a beam width of a water maser (S3 from 
\citealt{baudry96}).  This source is the brightest maser feature in the 
M82 starburst and is significantly brighter (by a factor of more than 
four) at 1667\,MHz than it is at 1665\,MHz, the most extreme line ratio 
out of all the detections.

\noindent{\bf 50.02+45.8} is located between a supernova remnant 
(41.29+59.7) and a possible water maser (P2 from \citealt{baudry96}) and 
is about a beam width from each.  It is only detected in the 1995 data-set 
where it is brightest at 1667\,MHz.

\noindent{\bf 49.71+44.4} is coincident with a possible water maser 
feature (P1 from \citealt{baudry96}) and also within a beam width of two 
H{\sc ii} regions (40.95+58.8 and 40.96+57.9).  It is brighter at 
1667\,MHz in every data-set in which it is detected, not appearing above 
the noise in any data-set at 1665\,MHz.  This maser appears to consist of 
two closely-spaced velocity components.

\noindent{\bf 48.45+42.1} is also within a beam width of an H{\sc ii} 
region (39.68+55.6), close to a water maser feature (S4 from 
\citealt{baudry96}) and is consistently brighter at 1667\,MHz.


\section{Discussion}

\subsection{Line ratios}

For an optically thin cloud in local thermal equilibrium, the ratio of 
emission for the ground states (1667:1665:1612:1720) is 
9:5:1:1, so S$_{1665}$$>$S$_{1667}$ is impossible without excitation such 
as that provided by a maser pumping mechanism.  Table \ref{ratioTable} 
lists the ratios between the two main lines for each maser from the observations discussed 
here, with limits given (based on the 3$\sigma$ detection limit) where a 
detection was made in only one of the two main lines.

In the MERLIN 1995 observations, four of the six masers are detected in 
both main lines, while 53.11+48.0 is detected only at 1665\,MHz and 
49.71+44.4 is present only at 1667\,MHz.  In 1997, the two masers detected 
were seen only at 1667\,MHz.  In 2002 six of ten are seen in both lines, 
53.11+48.0, 52.71+45.8 and 51.87+48.3 are seen only at 1665\,MHz, while 
49.71+44.4 is again seen only at 1667\,MHz.

\begin{table}
\begin{center}
\begin{tabular}{|c|ccc|c|}
\hline ID	& R$_{1995}$	& R$_{2002}$	& Closest feature\\ 
\hline
53.62+50.1 & -		& 1.02		& 44.93+63.9 (H{\sc ii})\\
53.11+48.0 & $<$0.47	& $<$0.41	& 44.43+61.8 (SNR)	\\
52.71+45.8 & -		& $<$0.15	& 44.01+59.6 (SNR)	\\
51.87+48.3 & -		& $<$0.45	& -			\\
51.23+44.5 & -		& 1.22		& 42.48+58.4 (H{\sc ii})\\
50.97+45.3 & 0.94	& 1.42		& 42.21+59.2 (H{\sc ii})\\
50.51+45.6 & -		& 0.69		& -			\\
50.36+44.2 & 4.27	& 4.72		& 41.64+57.9 (H{\sc ii})\\
50.02+45.8 & 1.18	& -		& 41.29+59.7 (SNR)	\\
49.71+44.4 & $>$2.03	& $>$3.04	& 40.96+57.9 (H{\sc ii})\\
48.45+42.1 & 1.67	& 2.26		& 39.68+55.6 (H{\sc ii})\\
\hline
\end{tabular} 
\caption[Ratio of S$_{1667}$:S$_{1665}$]{\label{ratioTable}Ratios of the 
measured intensities in the 1667\,MHz line to the 1665\,MHz line for each 
maser in M82.  Limits are given using the measured noise level in each 
cube.  The listed associations are the nearest continuum features 
(within a beam width of the maser position) from \protect\cite{mcdonald02}, 
labelled according to their B1950 coordinates.}
\end{center}
\end{table}

Of those which appear to be associated with H{\sc ii} regions, most are 
brighter at 1667\,MHz than at 1665\,MHz.  The only exception to this is 
50.97+45.3 which is reversed only in the 1995 observations.  The 
difference in magnitude between the two lines is, however, less than 
1$\sigma$.  Two of the masers, 53.11+48.0 and 52.71+45.8, appear to be 
associated with supernova remnants.  These two masers are both detected 
only at 1665\,MHz, if they are detected at all, with no emission apparent 
at 1667\,MHz.  Further discussion of apparent associations with other 
known features are given in the next section.

Investigating whether the masers vary in intensity between the epochs is 
not as simple as comparing their flux measurements.  The different channel 
widths used in each observation will determine, to some extent, the flux 
densities measured.  An alternative measure is the ratio between the two 
main line intensities in which the difference in channel width will have 
the same effect on each of the two lines.

It can be seen from Table \ref{ratioTable} that, to within the 
uncertainties of the flux measurements, all of the ratios are consistent 
over the different epochs of observation.  The limits on the 
ratio for 49.71+44.4 also vary by an amount larger than the uncertainties 
in the measured flux densities, with the 1667\,MHz line becoming 
comparatively bright in the 2002 data.  This source is interesting as it 
has additional structure which is apparent in the 2002 observations (see Section 
\ref{section_velocity}).

One source, 50.02+45.8, was detected in 1995, but is not visible in the 
2002 observations.  This source is one of the weakest seen in 1995 
and may have faded below the noise level in subsequent observations.
Several sources appear in the 2002 data-set, but not in the 1995 
data.  One likely cause of this is due to the width of the channels in the 
1995 data-set which will average narrow emission so that it appears 
weaker, possibly reducing it below the detection threshold.
Variability is often observed in Galactic masers (e.g. \citealt{caswell80}), 
so it is possible that these masers are varying in intensity between 
observations.

\subsection{Associations}

\begin{figure*} \centering
\includegraphics[width=10cm,angle=-90]{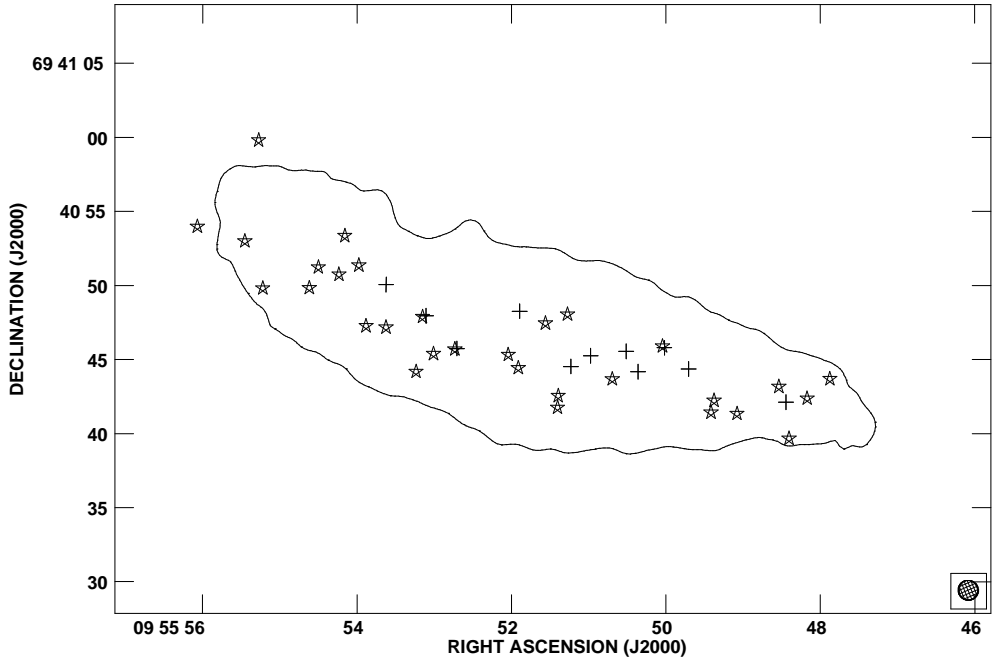}
\caption[Associations between main line masers and 
SNRs]{\label{figMasersSNR}The location of the main line masers (crosses) compared to the 
supernova remnants from \protect\cite{mcdonald02} (stars).  The outline of the galaxy is 
provided by a 1$-\sigma$ contour of the continuum emission at 18\,cm.}
\end{figure*}

\begin{figure*} \centering
\includegraphics[width=10cm,angle=-90]{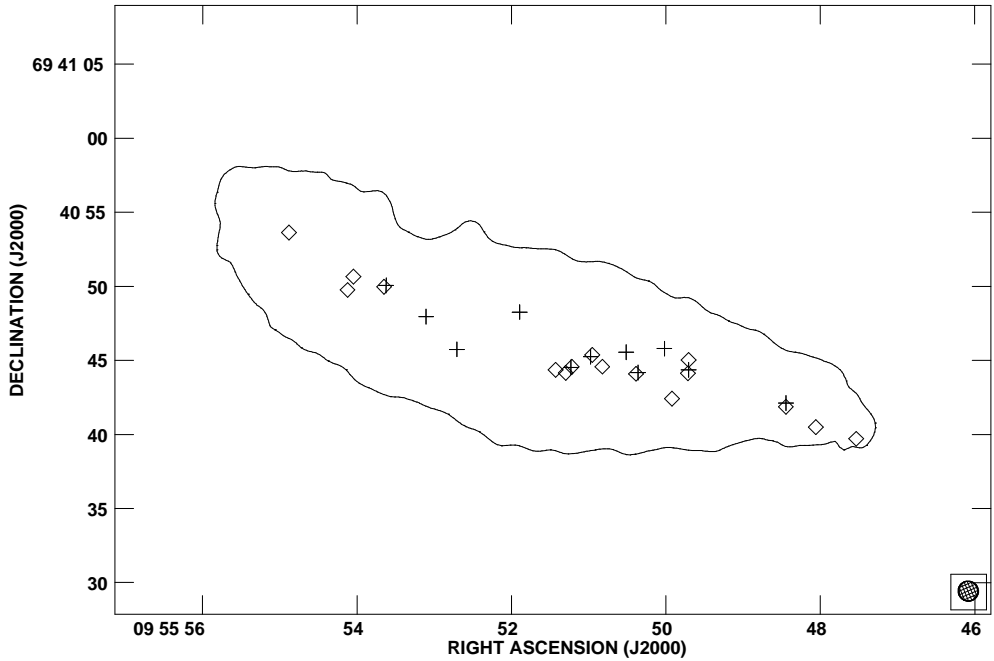}
\caption[Associations between main line masers and 
H{\sc ii} regions]{\label{figMasersHii}The location of the main line 
masers (crosses) compared to the H{\sc ii} regions from 
\protect\cite{mcdonald02} (diamonds).  The outline of the galaxy is provided by a 
1$-\sigma$ contour of the continuum emission at 18\,cm.}
\end{figure*}

\begin{figure*} \centering
\includegraphics[width=10cm,angle=-90]{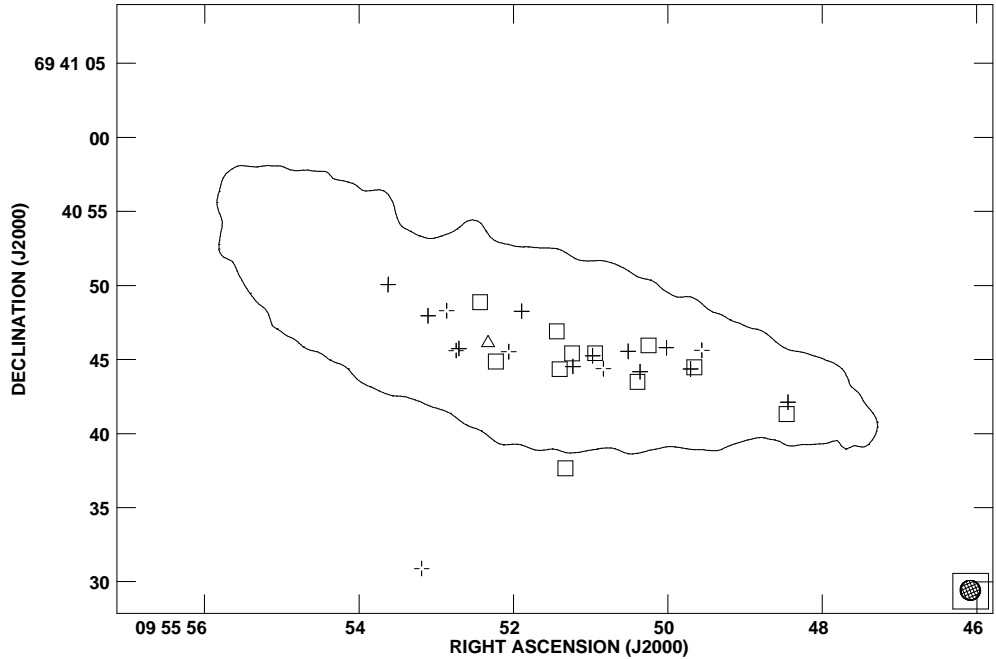}
\caption[Associations between main line masers and other 
masers]{\label{figMasers+}The location of the main line masers (crosses) compared to 
satellite line masers from \protect\cite{seaquist97} 
(broken crosses), water masers from \protect\cite{baudry96} (squares) and the 2.2 $\mu$m 
peak from \protect\cite{lester90} (triangle).  The outline of the galaxy is provided by a 
1$-\sigma$ contour of the continuum emission at 18\,cm.}
\end{figure*}

Figure \ref{figMasersSNR} shows the positions of both the main line maser 
detections described here and the SNRs from \cite{mcdonald02}.  
Figure \ref{figMasersHii} shows the positions of the OH masers along with 
the H{\sc ii} regions from \cite{mcdonald02}.  Figure \ref{figMasers+} 
shows the positions of the main line OH masers described here along with 
the satellite line masers detected by \cite{seaquist97}, the H$_2$O masers 
(both detections and possible detections) listed in \cite{baudry96} and 
the 2.2\,$\mu$m peak from \cite{lester90}.  In all three of these figures, 
the outline of M82 is represented by a 1$\sigma$ contour from the 
continuum emission in the VLA 2002 observation.  From these figures is can 
be  seen that most of the masers described here are closely associated 
with other known objects within M82.

Four sources (51.23+44.5, 50.97+45.3, 50.37+44.1 and 
48.45+42.1) are all apparently coincident with a water maser feature to 
within the size of the beam in the VLA 2002 observations.  A further two (50.51+45.6 and 
49.71+44.4) are located close to possible H$_2$O maser detections.  Of these 
associations, 50.51+45.6 is the most marginal as the separation is of the 
order of the beam size.  51.23+44.5 is within a beam of two separate water 
maser features, one of which is a marginal detection, listed in 
\cite{baudry96}.

Only a minority of the main line OH masers described here (four of eleven) 
do not appear coincident with water maser detections.  Of these, all 
except 51.87+48.3 are coincident with other known features.  All of the 
definite water maser detections listed by \cite{baudry96} are coincident 
with an OH main line maser detected in these observations.

Six OH masers (53.62+50.1, 51.23+44.5, 50.97+45.3, 50.36+44.2, 49.71+44.4 
and 48.45+42.1) appear to be associated with H{\sc ii} regions, while three 
(53.11+48.0, 52.71+45.8 and 50.02+45.8) are within a beam of known supernova remnants.
The large number of apparent associations is not surprising given the crowded fields and 
that maser emission is the result of an amplification of background continuum 
emission.
From the line ratios listed in the previous section it can be seen that 
the sources apparently associated with H{\sc ii} regions are generally 
brighter at 1667 than at 1665\,MHz, while those close to supernova 
remnants are brighter at 1665\,MHz (apart from 50.02+45.8).  In studies of 
Galactic masers such as 
that carried out by \cite{caswell98}, most masers in star forming regions 
are found to be brighter at 1665\,MHz than at 1667\,MHz.  This is 
discussed further in Section \ref{section_surveys}.

\begin{figure}
\centering
\includegraphics[width=7cm]{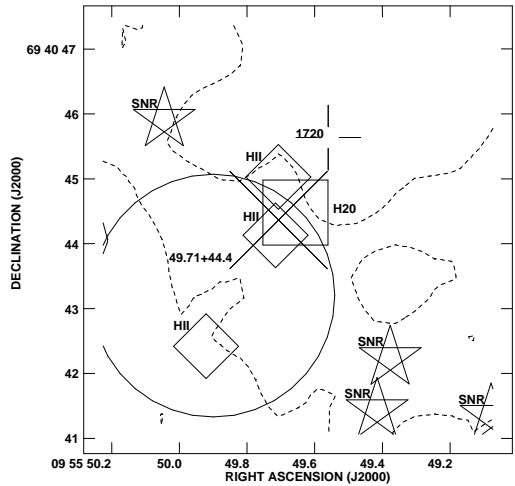}
\caption[The region around 49.71+44.4]{\label{fig4971}The region around 
49.71+44.4 showing the OH absorption from the VLA 2002 observation (dotted 
lines), the H{\sc i} shell (solid circle), H{\sc ii} regions (diamonds), 
water maser (square), supernovae (stars), 1720\,MHz maser (broken plus) 
and the OH maser position from the 2002 data-set (cross).  The symbols are 
1" in size, apart from that for 49.71+44.4 which is the size of the 
restoring beam.}
\end{figure}

The region around 49.71+44.4 contains two H{\sc ii} regions as well as a 
water maser.  The two H{\sc ii} regions are located at R.A. = 
49\rasec707, Dec. = 45\farcs03 and R.A. = 49\rasec716, Dec. = 
44\farcs13.  Along with the multiple velocity components seen in this 
maser in the 2002 observations, this region of M82 is 
particularly interesting.  Figure \ref{fig4971} shows this region in more 
detail.  The location of this maser also appears to be on the edge of one 
of the H{\sc i}/CO shells (\citealt{wills02}) and there is a 1720\,MHz maser detection 
within a beam width.

The brightest X-ray source in M82 is located at 50\rasec2, 46\farcs7 
(\citealt{kaaret01}) with a 1$\sigma$ error circle of 0.7 arcseconds.  
None of the maser detections lie within this circle, the closest detection 
being 50.02+45.8 which lies at a distance of 2\farcs85 (a linear distance 
of 44\,parsecs) so is clearly not associated.

The black hole candidate described by \cite{ptak99} is located at R.A. = 
09$^{\rm h}$55$^{\rm m}$52$^{\rm s}$, Dec. = +69\degr40'48" with a 1.3 
arcminute diameter error circle.  The nearest maser detection to this 
source is 51.87+48.3, a source only present in the 2002 data-set at 
1665\,MHz.  The separation between these positions is 1\farcs9 so 
considering the uncertainties it is possible that they could be 
associated.

The AGN candidate suggested by \cite{wills97} and \cite{seaquist97} is 
spatially coincident with the maser 52.71+45.8.  This maser is apparently associated 
with an SNR, and one of only three detected at 1665\,MHz while not 
detected at all at 1667\,MHz.  It is also the only one coincident with a 
satellite maser from \cite{seaquist97}, a position which displayed 
absorption at 1720\,MHz, but strong emission at 1612\,MHz.

\begin{figure*} \centering
\includegraphics[width=16cm]{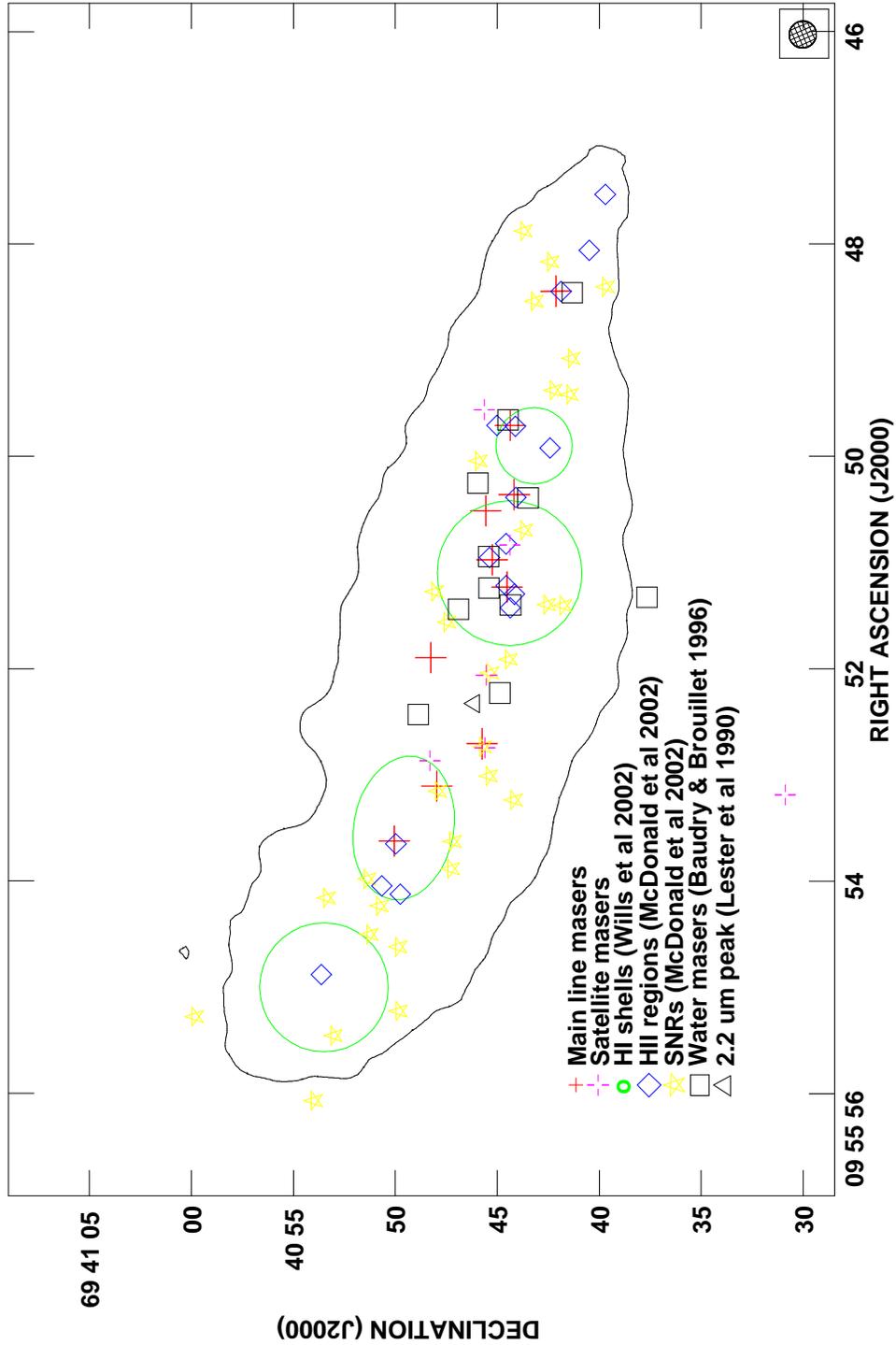}
\caption[The NastyFigure]{\label{figMasersEtc}The location of the main 
line masers in M82 with other features shown for comparison.  The solid 
line is a 1$\sigma$ contour of the continuum emission from the VLA data 
cube.  The full crosses mark the positions of the main line masers 
reported here.  The crosses with gaps mark the 1612/1720\,MHz masers 
reported in \protect\cite{seaquist97}.  The diamonds and stars mark the 
H{\sc ii} regions and SNRs from \protect\cite{mcdonald02} respectively.}  
\label{fig_all}
\end{figure*}

For completeness, Figure \ref{figMasersEtc} shows the positions of the 
masers listed in Table \ref{tinymaserTable} along with the positions of 
the H{\sc ii} regions and supernova remnants listed in \cite{mcdonald02}, 
the satellite masers listed in \cite{seaquist97}, the H{\sc i} shells 
listed in \cite{wills02}, H$_2$O masers from \cite{baudry96}, and the 
2.2\,$\mu$m peak from \cite{lester90}.


\subsection{Comparison with other maser detections}

The two brightest masers, 50.97+45.3 and 50.36+44.2, are coincident with
the OH masers reported by \cite{weliachew84}, corresponding to their
masers m2 and m1 respectively.  In their data, one source (labelled m1) 
is seen at 1665\,MHz, while both their m1 and m2 are detected at 
1667\,MHz.  In the 2002 VLA data, both masers are detected in both main 
lines at velocities of 80 and 98 $\pm$ 18 \kms\ respectively.  This is 
consistent with the velocities of the masers in \cite{weliachew84} where 
their m1 is at 65 to 80 \kms, and m2 has a velocity of 85 to 100 \kms.

Two of the OH masers reported here are apparently coincident with the H$_{2}$O 
features reported by \cite{baudry96} to within the size of the 1\farcs4 VLA beam 
in their observations.  Their S2 is 0\farcs29 from the MERLIN position of 50.97+45.3 while 
the brightest water maser, S3, lies 0\farcs63 from the MERLIN position of 50.36+44.2.  S1 and S4 from \cite{baudry96}  
are both $>$1\farcs6 from the nearest MERLIN detection.  One other OH maser, 49.71+44.4, is 
located 0\farcs57 from a possible water maser detection listed by the same 
authors.
The velocities of the H$_2$O masers are consistent (within the 
uncertainties) with those measured for the OH detections presented here, 
apart from perhaps 50.36+44.2/S3 although even here the difference is 
small: \cite{baudry96} measure 103.9\,\kms for S3 where as 50.36+44.2 is 
found to have a velocity of 80$\pm$18\,\kms from the 2002 VLA data.

Only one OH maser is coincident with a satellite line maser position 
reported by \cite{seaquist97}: the separation of 52.71+45.8 with feature 4 
from that paper (a feature seen in absorption at 1612\,MHz but strong emission 
at 1720\,MHz) is only 0\farcs57.  Both observations were carried out 
using the VLA with a $\theta_{\rm HPBW} \sim$ 1\farcs4.  This position is 
also coincident with the SNR 44.01+59.6, one of three definite 
detections in this sample coincident with an SNR.  Feature 4 in 
\cite{seaquist97} had a measured flux of 31.6$\pm$1.7\,mJy at 1612\,MHz in 
1996 November and a peak of only 1.98\,mJy\,beam$^{-1}$ at 1665\,MHz in 
2002 November.  At 1720\,MHz there is no emission apparent, only an 
absorption feature is present at this position.


\subsection{Spatial structure}

Most of the masers are unresolved in the observations detailed here.  This 
is not surprising given that the typical size of Galactic maser spots is 
much less than a parsec, considerably less than the resolution of the 
observations detailed here.  One maser feature does show extension in the 
VLA data which resolves into two components when observed with MERLIN.

\begin{figure}
\centering
\includegraphics[width=6cm,angle=-90]{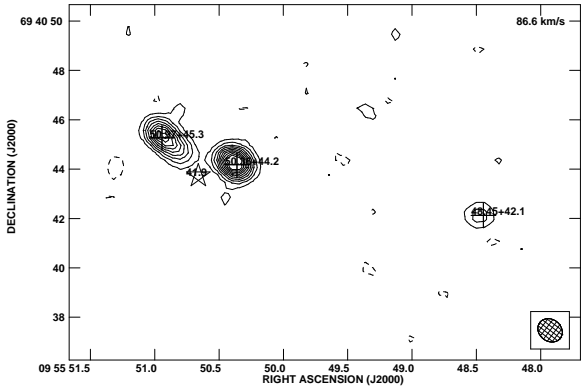}
\caption[Central masers as seen in the 2002 VLA data]{Three of the central 
masers in a continuum-subtracted single channel from the VLA 2002 data-set showing the 
extension of maser 50.97+45.3.  The position of the brightest continuum 
feature (the SNR 41.9+57.5) is marked by a star.  Contours are at (-1, 1, 
2, 3, 4, 5, 6, 7, 8, 9, 10) $\times$ 1.751 mJy/beam.}
\label{fig_extended}
\end{figure}

\begin{figure}
\centering
\includegraphics[width=6cm]{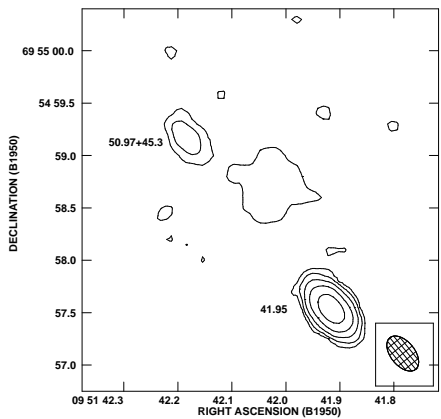}
\caption[Central masers as seen in the 1995 MERLIN data]{One of the 
central masers in a single channel from the MERLIN 1995 data-set showing 
the extension of maser 50.97+45.3.  This image shows both the maser and continuum features.  
The position of the brightest 
continuum feature (the SNR 41.9+57.5) is toward the bottom of the image.  
Contours are at (-1, 1, 2, 4, 8, 16, 32 and 64) $\times$ 2.6\,mJy/beam.}
\label{fig_extended_merlin}
\end{figure}

The extension to the SW of 50.97+45.3 in the VLA 2002 data-set can be seen 
in Figure \ref{fig_extended}.  There is also a marginally detected second 
spatial component to this maser in the MERLIN 1995 data at the same 
position angle as the elongation seen in the 2002 data (see Figure 
\ref{fig_extended_merlin}), although it is only present at just over 
3$\sigma$.  The weak detection suggests that there is structure in the 
MERLIN data-set on scales larger than those to which the interferometer is 
sensitive which is resolved out.  Using the AIPS task {\sc jmfit} to fit 
Gaussian components to a region including masers 50.97+45.3 and 50.36+44.2 
in the VLA 2002 data produces a better quality fit when three components 
are used, rather than two.  The positions of the resulting fitted Gaussian 
components are given in Table \ref{table_gaussians}.

\begin{table}
\begin{center}
\begin{tabular}{|cc|c|}
\hline
RA		& Dec		& S (mJy)	\\
\hline
50\rasec95	& 45\farcs417	& 14.8		\\
50.81		& 44.802	& 7.08		\\
50.38		& 44.306	& 24.4		\\
\hline
\end{tabular} 
\caption[Fit components for 50.97+45.3]{\label{table_gaussians}
List of the Gaussian components fitted to 50.97+45.3 and 50.36+44.2 for 
channel 47 in the VLA 2002 data.  Each component had the same major and 
minor axis, and position angle as the restoring beam (1.34$\times$1.31 at 
23\degr.19).}
\end{center}
\end{table}

The maser 49.71+44.4 appears to consist of multiple components at 
different velocities in the 1667\,MHz line.  This velocity structure is 
seen in the MERLIN 1995 and VLA 2002 data and may consist of 
multiple spatial components within a small region.  This source is 
discussed in more detail in Section \ref{section_velocity}.


\subsection{\label{section_velocity}Velocity distribution}

Figure \ref{fig_2002spectra} shows the spectra of the ten definite masers 
detected at greater than 5$\sigma$ in the 2002 observations.  Most of the 
masers are clearly visible above the absorption but it can be seen that 
others, like 49.71+44.4, may be hidden in the absorption.  It can also be 
seen that while some of the masers have the same line-of-sight velocity as 
the underlying absorption, several are blue or red shifted slightly.  The 
velocities of each feature measured in this data-set are given in Table 
\ref{tinymaserTable}.

\begin{table}
\begin{center}
\begin{tabular}{|c|c|c|}
\hline ID & Velocity & Nearest continuum\\
(J2000)  & (\kms) & feature (B1950)\\
\hline
53.62+50.1 & 347$\pm$9	& 44.93+63.9 (H{\sc ii}) \\
53.11+48.0 & 281$\pm$9	& 44.43+61.8 (SNR)	\\
52.71+45.8 & 233$\pm$9 	& 44.01+59.6 (SNR)  \\
51.87+48.3 & 238$\pm$9	& -	\\
51.23+44.5 & 119$\pm$9	& 42.48+58.4 (H{\sc ii}) \\
50.97+45.3 & 96$\pm$9	& 42.21+59.2 (H{\sc ii}) \\
50.51+45.6 & 203$\pm$9	& -	\\
50.36+44.2 & 79$\pm$9	& 41.64+57.9 (H{\sc ii}) \\
49.71+44.4 & 148$\pm$9	& 40.96+57.9 (H{\sc ii}) \\
48.45+42.1 & 107$\pm$9	& 39.68+55.6 (H{\sc ii}) \\
\hline
\end{tabular} 
\caption[Definite maser detections in M82]{\label{tinymaserTable}The velocities of the ten
definite maser detections in the VLA 2002 data.  The features are named according to their
J2000 positions (relative to 09$^{\rm h}$55$^{\rm m}$ +69\degr40').  
Velocities are measured with respect to the line rest velocity and where a
maser is detected in both lines the velocities are consistent between the
two.  The nearest continuum features are from, and are labelled according
to, the convention used in \protect\cite{mcdonald02} (their corresponding J2000 IDs are given in Table 
\protect\ref{table_allmasers}.}
\end{center}
\end{table}

\begin{figure*}
\centering   
\includegraphics[width=10cm]{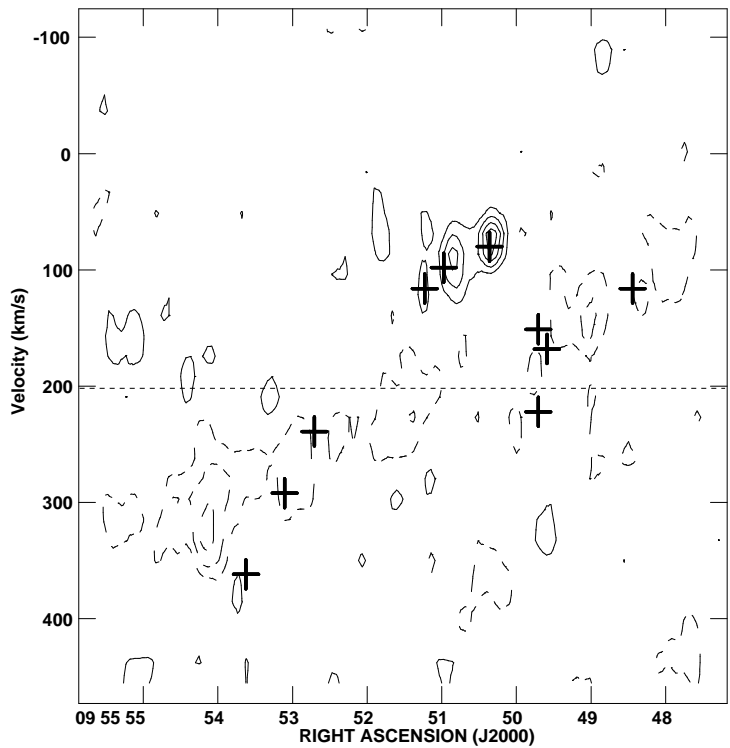}
\caption[Comparison of masers in the 2002 data set]{\label{figPV}A comparison between the 
maser velocities and the absorption.  Crosses represent measurements of 
maser velocities from the 2002 data.  The contours represent the OH 
absorption (and emission) from the VLA 2002 dataset and the horizontal dashed line 
represents the systemic velocity of M82.}
\end{figure*}

\begin{figure*}
\centering
\psfig{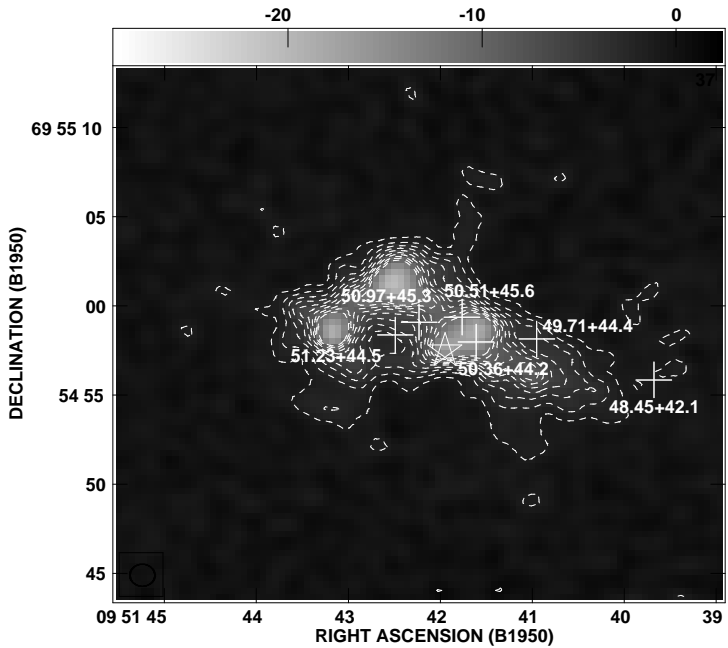}
\caption[Masers in relation to the H{\sc i} absorption]{The positions of 
the masers in the centre of M82 in relation to the H{\sc i} absorption 
shell seen by \protect\cite{wills02} at a velocity of 173\,\kms.  
Contours are of the H{\sc i} absorption at (-10 to +10 in steps of 1) 
$\times$ 1.1\,mJy/beam.}
\label{fig_masers+h1abs}
\end{figure*}

The velocities measured for most of the masers lie on a line consistent 
with the rotation of a disk; this can be seen in Figure \ref{figPV}.  
The sources 53.62+50.1, 53.11+48.0, 52.71+45.8, 51.87+48.3 and 48.45+42.1 
are at the same velocity as the OH absorption from the galaxy.  
On the other hand, 51.23+44.5, 50.97+45.3, 50.36+44.2 and 49.71+44.4 lie 
on an arc which is blue shifted with respect to the absorption at the same 
position.  The location of this arc is coincident with the possible shell 
seen in CO emission by \cite{matsushita00} and H{\sc i} absorption by 
\cite{wills02} (see Figure \ref{fig_masers+h1abs}) and could be associated 
with a superbubble
around the unusual source 41.95+57.5 (\citealt{muxlow05}).  
Figure \ref{fig_masers+ohabs} shows that there is little OH absorption 
around this position.   Another possibility is that these features, rather 
than being associated with a bubble or shell-like feature, are situated on 
the line of the inner orbits of the possible bar structure described by 
\cite{wills00}.

One of the masers associated with this velocity feature, 49.71+44.4 is 
unusual.  In the VLA 2002 data-set there is an emission feature at 
2.34 mJy in the 1667\,MHz line with a velocity of 151\,\kms.  This feature 
has a weaker counterpart in the 1665\,MHz line with a peak of $<$1\,mJy at 
a velocity of 168\,\kms.  There also appears to be a possible very weak 
second emission feature in the 1667\,MHz line with a velocity of 
222\,\kms.  This second feature is comparable with the noise in the map so 
was initially ignored.  However, this second component in the 1667\,MHz 
line is also present in the MERLIN 1995 data.


\section{Comparison with other surveys}\label{section_surveys}

\cite{ball68} compare OH maser detections at various frequencies with 
other objects.  They find that H{\sc ii} regions are generally brightest 
in either of the main lines (class I in the terminology of 
\citealt{turner69}) while those associated with SNRs or other non-thermal 
sources associated with H{\sc ii} regions are brightest at 1612 or 
1720\,MHz (Turner's class II) and suggest that the processes in 
non-thermal sources could inhibit the main line emission.

In their conclusions, \cite{pavlakis96b} state that where both 1665 
and 1667\,MHz lines are spatially coincident, an FIR radiation field 
(assumed to be due to thermal emission from dust) must be present.  They 
also state that the 1665\,MHz line is typically stronger than that at 
1667\,MHz.  Most of the masers seen M82 appear to display the opposite 
behaviour, with the 1665 emission weaker, in most cases, than that at 
1667\,MHz.

A survey by \cite{caswell98} of OH ground state masers in the Southern 
Galactic plane confirmed previous conclusions that the 1665 and 1667\,MHz 
lines often occur together in space.  The 1665\,MHz line is usually 
stronger than the 1667\,MHz line by a factor of around 3, although the 
1667\,MHz line is stronger in some sources by up to a factor of 4.

\cite{caswell80} survey a selection of the Galactic plane for OH in a 
restricted longitude range.  They detect a total of 40 main line maser 
sources, most of which are brighter at 1665 than 1667\,MHz.  Most are also 
associated with known H{\sc ii} regions, while two of the eight remaining 
sources with no known H{\sc ii} associations may be associated with SNRs.  
Three of the masers in M82 are similarly associated with SNRs rather than 
H{\sc ii} regions, although the possibility that there are undetected 
H{\sc ii} regions along the same line of sight cannot be discounted.

\cite{wilson72} (and later \citealt{slootmaker85}) searched for OH masers 
associated with late-type stars in the Milky Way and found that most 
sources displaying emission in the main lines were brighter at 1667\,MHz 
than at 1665\,MHz.  They find that a positive 1667:1665 brightness 
ratio appears to be more characteristic of evolved stars, rather than star 
formation and H{\sc ii} regions, with the differences in strength of 
emission due to the properties of the dust grains in the masing cloud 
(\citealt{elitzur92}).

The most surprising result to come from this study is that, contrary to the 
results of other surveys discussed here, most of the masers in M82 are 
brighter at 1667\,MHz and appear associated with H{\sc ii} regions.  This 
would imply that most of these masers are associated with regions of 
evolved stars, rather than H{\sc ii}/OH regions as they appear to 
be.  However, the apparent associations here are not necessarily physical and could 
just be due to line of sight coincidences since maser emission is the 
result of the amplification of background continuum emission.  Another 
possibility is that, as the resolution of these observations is 3\,pc at 
best, they are not physically associated with the observed H{\sc ii} 
regions but with nearby regions of evolved stars.  Higher angular ressolution 
measurements would be required in order to investigate whether the 
associations are real and physical, or just due to line-of-sight effects.

\section{Conclusions}

Nine new OH main line masers have been detected in the central starburst 
region of the nearby galaxy M82.  On a position-velocity plot of the 
galaxy, most of the masers lie along the same axis as the gas.  Four of 
the masers are blue-shifted from the main distribution and could possibly 
be on the edge of an expanding shell or caused by gas orbiting within a bar.  Most appear 
associated with H{\sc ii} regions but are more 
characteristic of masers associated with late-type stars.

Due to the spectral resolution of the 2002 VLA observations the masers 
were not resolved in frequency, with most only visible in two channels.  As 
the individual spots are an order of magnitude more luminous than typical 
Galactic masers, it is likely that they are made up of more than one 
individual masing region, so would have structure both spatially and in 
frequency.  In order to investigate this further, observations have recently been made using 
the European VLBI Network in order to provide higher spatial resolution (much less than 
1\,pc at the distance of M82), and the VLA at higher spectral resolution.
It is hoped that further investigation using these two observations will show evidence 
of both spatial and velocity structure.

It is likely that there are more main line OH masers in M82 but, due to 
the depth of absorption and the low velocity resolution, faint or narrow 
masers could be buried to the extent that they are undetectable in the 
2002 observations.  The recent high spectral resolution VLA observations will have a gain 
in sensitivity over the 2002 observations which should enable 
the detection of weaker masers.

Most of the maser spots detected are unresolved and likely consist of 
large numbers of small masing clouds.  Any variability could be due either 
to variability of a few strong sources in a particular spot, the 
creation/destruction of many weaker maser sources within a spot, or a 
combination of both.

The association of most maser features with H{\sc ii} regions is not 
surprising given the results of surveys carried out in out own Galaxy, 
although the comparative strength of the 1667\,MHz line is more 
characteristic of late-type stars than H{\sc ii} regions.  The association 
of two masers with known supernova remnants, while unusual, is not unique.  
A more sensitive survey for emission at 1612 and 1720\,MHz would also help 
determine the nature of the maser sources.

\begin{table*}
\centering
\begin{tabular}{|c|cc|cc|cc|cc|cc|}
\hline
ID	& \multicolumn{2}{c}{1995}	& \multicolumn{2}{c}{1997}	& \multicolumn{2}{c}{2002}	& Nearest feature	& Nearest feature \\
(J2000)	& S$_{1665}$	& S$_{1667}$	& S$_{1665}$	& S$_{1667}$	& S$_{1665}$	& S$_{1667}$	& (B1950) 		& (J2000)	  \\ \hline
53.62+50.1	& $<$2.6 & $<$2.6 & $<$17	& $<$18	& 2.44		& 2.49		& 44.93+63.9 (H{\sc ii})	& 53.65+50.0	\\	
53.11+48.0	& 5.57	 & $<$2.6 & $<$17	& $<$18	& 1.86		& $<$0.77	& 44.43+61.8 (SNR) 		& 53.14+47.8	\\	
52.71+45.8	& $<$2.6 & $<$2.6 & $<$17	& $<$18	& 5.02		& $<$0.77	& 44.01+59.6 (SNR) 		& 52.73+45.7	\\	
51.87+48.3	& $<$2.6 & $<$2.6 & $<$17	& $<$18 & 1.71		& $<$0.77	& -				& -		\\
51.23+44.5	& $<$2.6 & $<$2.6 & $<$17	& $<$18	& 4.85		& 5.93		& 42.48+58.4 (H{\sc ii})	& 51.22+44.5	\\	
50.97+45.3	& 8.35	 & 7.83	  & $<$17	& 23.0	& 13.0		& 18.5		& 42.21+59.2 (H{\sc ii})	& 50.95+45.2	\\	 
50.51+45.6	& $<$2.6 & $<$2.6 & $<$17	& $<$18	& 1.85		& 1.28		& -				& -		\\	
50.36+44.2	& 7.01	 & 29.9	  & $<$17	& 57.2	& 10.7		& 50.5		& 41.64+57.9 (H{\sc ii})	& 50.39+44.1	\\	
50.02+45.8	& 4.30	 & 5.09	  & $<$17	& $<$18	& $<$0.77	& $<$0.77	& 41.29+59.7 (SNR)		& 50.06+45.9	\\	 
49.71+44.4	& $<$2.6 & 5.29	  & $<$17	& $<$18	& $<$0.77	& 2.34		& 40.96+57.9? (H{\sc ii})	& 49.71+44.1	\\	 
48.45+42.1	& 4.15	 & 6.94	  & $<$17	& $<$18	& 2.64		& 5.97		& 39.68+55.6 (H{\sc ii})	& 48.44+41.9	\\	
\hline
\end{tabular}
\caption[List of all masers and their fluxes]{\label{table_allmasers}
All masers detected in the data-sets discussed here.  Fluxes 
are given for each masers in each line in which is was detected in each 
epoch, along with the closest continuum feature from 
\protect\cite{mcdonald02} given in both B1950 and J2000 coordinates.}
\end{table*}



\subsection*{Acknowledgements}

MERLIN is run by the University of Manchester as a National Facility on behalf of PPARC.
The VLA is operated by the National Radio Astronomy Observatory, a facility of the National Science Foundation operated under cooperative agreement by Associated Universities, Inc.
MKA acknowledges support from a PPARC studentship.
RJB acknowledges support from the European Commission's I3 programme "RADIONET" under contract 505818.

\bibliographystyle{mn}
\bibliography{refs}

\end{document}